# The Informational Model of the Holobiont: Statistical Tests for Selection and Extension to a Theory of Variable Interactions

Antonio Carvajal-Rodríguez[1*]

[1]Centro de Investigación Mariña (CIM), Departamento de Bioquímica, Genética e Inmunología. Universidade de Vigo, 36310 Vigo, Spain

*Correspondence address: Centro de Investigación Mariña (CIM), Departamento de Bioquímica, Genética e Inmunología. Universidade de Vigo, 36310 Vigo, Spain. Tel: +34 986130052. E-mail: acraaj@uvigo.es

## Abstract

We review, clarify, and generalize a recently proposed evolutionary information-theoretic model of the holobiont, in which evolutionary change is quantified using Jeffreys divergence and partitioned into contributions from the host, microbial components, and host-microbiome associations. Building on these partitions, we develop statistical tests to distinguish informational changes in host and microbial marginal distributions from changes in host-microbiome associations and conditional dependence among microbial components within host types. We then extend the framework to multicomponent groups subject to within- and between-group selection and, ultimately, to a general hierarchical formulation that we call the Theory of Variable Interactions (TVI). In this formulation, biological units may contain interacting components and may themselves form higher-level sets, allowing informational change to be partitioned recursively into marginal and association components across an arbitrary number of organizational levels. The framework encompasses previously studied models of the tragedy of the commons and of aggregate and multicomponent holobiont selection, and, as a further specialization, informational models of non-random mating and sexual selection. TVI therefore provides a unified information-theoretic framework for quantifying evolutionary change in both biological entities and their interactions, with local statistical dimensionality preserved across hierarchical levels while higher levels introduce additional relational dimensions.

## 1 Introduction

A holobiont is a composite biological system consisting of a host and the ecological community of microorganisms that constitutes its microbiome (Margulis and Fester, 1991; Roughgarden, 2020). The combined genomes of the host and its associated microorganisms form the so-called hologenome (Theis et al., 2016; Zilber-Rosenberg and Rosenberg, 2008). The hologenotype represents a particular configuration of the hologenome in an individual holobiont, whereas its phenotypic expression, the holophenotype, encompasses the physiological, morphological, and behavioural properties emerging from the host-microbiome system. Holobiont selection can therefore be understood as differential survival or reproduction among holobionts according to their holophenotypes (Roughgarden, 2020), thereby changing the relative frequencies of host, microbial, and joint host-microbiome configurations.

Recent developments in holobiont biology have emphasized the need for quantitative hologenomic approaches and for statistical methods and analytical tools capable of disentangling host, microbial, and intergenomic contributions to holobiont variation (Bordenstein and The Holobiont Biology Network, 2024). In this direction, Week et al. (2025) extended quantitative genetics to model the combined contribution of host genetic and microbial factors to phenotypic variation, partitioning additive trait variation into host-genetic, microbial, and gene-microbe covariance components and allowing explicit gene-microbe interactions. The informational model developed here provides a complementary decomposition of host-microbiome evolution. Whereas the quantitative-genetic approach partitions variation in a host trait according to the factors contributing to that variation, the informational approach partitions the realized change produced by selection into changes in host and microbial marginal distributions and changes in their association structure. Thus, the two frameworks address related but distinct aspects of host-microbiome evolution: the sources of phenotypic variation and the distributional information generated by evolutionary change.

The informational holobiont model developed below belongs to a broader mathematical framework in which associations, alongside their constituent entities, become explicit objects of evolutionary analysis. Biological units may be organized hierarchically: components form groups, and groups may themselves combine into higher-level sets. At each level, selection can alter both the marginal representation of the constituent classes and the frequencies with which particular combinations occur. A holobiont provides a direct example: its joint fitness depends on a particular combination of host and microbiome, whereas the apparent fitnesses of host or microbial classes arise as marginal quantities derived from these joint fitnesses. The resulting informational partition can therefore distinguish changes attributable to differential representation of hosts or microbial components from changes attributable specifically to host-microbiome associations.

Non-random mating provides a particularly simple specialization of the same structure. In this case, one sex occupies the formal role of the host and the other that of the microbiome, so that the fundamental unit is a female-male pairing rather than a host-microbiome combination. The relevant joint parameter is the mutual mating propensity or mutual mating fitness of a pair. Its marginal effects quantify differential mating success among types, that is, sexual selection, whereas deviations of pair frequencies from those expected from the marginals quantify mate choice or assortative mating in informational terms (Carvajal-Rodríguez, 2024, 2018). In both cases, the primary variable refers to an association, while quantities attributed to individual classes arise as marginals of the corresponding joint distribution.

In the simplest limit, the same mathematical structure reduces to a single-component system. With only one component, no association term can arise, and the model reduces to the standard replicator relation between changes in frequency and relative fitness. In this case, the selection-induced mean change in log relative fitness is quantified by the Jeffreys divergence, recovering the information-theoretic connection originally formulated by Frank (2012).

We refer to the general framework describing such hierarchically organized and potentially changing associations as the Theory of Variable Interactions (TVI). Here, interaction is used in a broad relational sense, encompassing the organization of biological entities within joint configurations, whereas association denotes its formal representation as statistical dependence in joint frequency distributions. Thus, the framework quantifies changes in association structure without requiring them to correspond to a particular mechanistic or causal interaction. In this sense, variable refers to the relational structure itself: associations among components may change in strength, structure, or informational contribution as a consequence of evolutionary processes. The same association-based logic can be applied recursively when groups themselves become components of higher-level sets, thereby extending the framework to within- and between-group selection and, more generally, to hierarchical multicomponent systems.

Within this framework, the present study revises and generalizes the informational model of the holobiont introduced by Carvajal-Rodríguez (2026). We develop statistical tests for determining whether the informational effects of selection are attributable to changes in the host, microbial components, or specific host-microbiome associations, and illustrate their application with a simple simulated example involving octopus holobionts. We then extend the same association-based structure to hierarchical multicomponent sets, yielding the recursive formulation of TVI.

# 2 The Selection Information Model of the Holobiont

It remains unclear whether the holobiont should be regarded as a single evolutionary unit, because the host and microbiome may retain partial evolutionary independence and microbial transmission is often horizontal rather than strictly vertical. Nevertheless, holobiont-level selection may still be evolutionarily relevant even when host-microbiome integration and transmission are incomplete (Roughgarden, 2020).

## 2.1 Fitness Model

We define a holobiont as comprising a host (*H*) and a microbial component (*M*), and decompose its fitness (*W*) into host, microbial, and association components:

$$W_{ij}(t)=W_H(i,t)\cdot W_M(j,t)\cdot a_{ij}(t)$$

where $W_H$ is the host contribution, $W_M$ is the microbial contribution and $a_{ij}$ is the host-microbiome association factor, accounting for non-independent effects between the two components. On the logarithmic scale:

$$logW_{ij} = logW_H(i) + \log W_M(j) + \log a_{ij},$$

where, for simplicity, the generation index ($t$) is omitted throughout.

This decomposition is the foundation for the informational partition developed below.

## 2.2 Informational Model of Selection in the Holobiont: An Aggregate Microbiome

The aggregate formulation reviewed in this section was introduced by Carvajal-Rodríguez (2026) and is revisited here, with some reformulation for clarity, as the basis for the statistical tests developed below. In this formulation, the microbiome is treated in aggregate, as a single component of the holobiont, without explicitly representing its internal structure or the contributions of its individual microbial components. The model is subsequently extended to a structured multicomponent microbiome.

### *2.2.1 Model Formulation and Informational Partition*

Let $p_i$ be the frequency of host type $i$ and $m_j$ the frequency of microbiome type $j$. Then, before selection, the expected frequency of holobionts of type $i \times j$ is

$$q_{ij} = p_i m_j.$$

The independence $q_{ij} = p_i m_j$ is the null reference of random assembly from the pools available to each holobiont; it is not a statement about the host population before selection. Transmission (horizontal, vertical or mixed) is part of $W$ and therefore of the association factor $a_{ij}$. For example, suppose that microbial type $k_1$ has frequency 0.1 in the source pool but is found in every host of type $i$ after selection. This represents a departure from the random-assembly reference, captured by an effective association factor (with $a_{i,k_1}$ large relative to $a_{i,k}$ for the remaining microbial types). An enrichment of $k_1$ in all host types is instead captured by the marginal term $J_M$. The reference frequency *m* must be the pool actually available at assembly: if only the global population frequency is used and holobionts share environments (tanks, sites), apparent associations may arise without any preference. The statistic $a'_{ij}$ measures the net association between assembly and subsequent survival and does not distinguish at which stage the preference acts unless the design allows it.

As we have already seen, the fitness of this holobiont resulting from combination of host *i* with microbiome *j* is defined as

$$W_{ij} = W_H(i) \times W_M(j) \times a_{ij} \quad \forall i, j : W_H(i) > 0, W_M(j) > 0, a_{ij} > 0,$$

and

$$\bar{W} = \sum_{i,j} p_i m_j W_{ij},$$

then the expected frequency of the holobiont after selection

$$q'_{ij} = p_i m_j \frac{W_{ij}}{\bar{W}}.$$

And the marginal frequencies after selection

$$p'_i = \sum_j q'_{ij} = p_i W_H(i) \frac{A_i}{\bar{W}},$$

$$m'_j = \sum_i q'_{ij} = m_j W_M(j) \frac{B_j}{\bar{W}},$$

where

$$A_i = \sum_j m_j W_M(j) a_{ij},$$

$$B_j = \sum_i p_i W_H(i) a_{ij},$$

and the post-selection association factor $a'_{ij}$ relative to the new marginal frequencies:

$$a'_{ij} = \frac{a_{ij} \cdot \bar{W}}{A_i B_j}.$$

The character ($Z$) associated with the holobiont is defined as the logarithm of its relative fitness:

$$Z = \log\left(W / \bar{W}\right),$$

then the average change of $Z$ due to selection is defined as (Frank, 2012)

$$J_T = \sum_{i,j} \left(q'_{ij} - q_{ij}\right) \log \frac{q'_{ij}}{q_{ij}},$$

where log denotes the natural logarithm.

The Jeffreys divergence, rather than the one-sided Kullback-Leibler divergence, is the quantity of interest because it is exactly the mean change of the character $Z$. Let

$$\omega_{ij} = \frac{q'_{ij}}{q_{ij}} = \frac{W_{ij}}{\bar{W}},$$

so that $\Sigma q_{ij}\omega_{ij} = 1$ and Z = log $\omega_{ij}$. The mean of $Z$ after selection minus its mean before selection is

$$J_T = \sum_{i,j} \left(q'_{ij} - q_{ij}\right) \log \omega_{ij} = \mathrm{E}_{q'}[Z] - \mathrm{E}_q[Z] = \mathrm{Cov}_q(\omega, \log \omega)$$

which is also the covariance between relative fitness and its logarithm (Carvajal-Rodríguez, 2024, 2018; Frank, 2012). The two one-sided divergences are the mean of Z after selection and minus the mean of Z before selection, respectively; $J$ is their sum.

And we obtain the partition

$$J_T = J_H + J_M + J_{assoc} + E_{holo},$$

where

$$J_H = \sum_i \left(p'_i - p_i\right) \log \frac{p'_i}{p_i} = \sum_i \left(p'_i - p_i\right) \log \frac{W_H(i) A_i}{\bar{W}},$$

$$J_M = \sum_j \left(m'_j - m_j\right) \log \frac{m'_j}{m_j} = \sum_j \left(m'_j - m_j\right) \log \frac{W_M(j) B_j}{\bar{W}},$$

$$J_{assoc}=\sum(q'-r)\log\frac{q'}{r}=\sum(q'-r)\log a'_{ij}=\sum_{i=1}^{K_H}\sum_{j=1}^{K_M}(q'_{ij}-r_{ij})\log\left(\frac{\bar{W}a_{ij}}{A_iB_j}\right),$$

where $K_H$ and $K_M$ are the numbers of host and microbiome classes, respectively and $r$ is the independence reference:

$$r_{i,j}=p'_i\cdot m'_j,$$

that is, the independent distribution matching the post-selection marginals of $q'$. And

$$E_{holo}=\sum(r-q)\log\frac{q'}{r}.$$

Selection changes the host and microbial marginal distributions through the effective marginal fitnesses $W_H(i)A_i$ and $W_M(j)B_j$, whereas changes in the host-microbiome association structure are captured by $J_{\text{assoc}}$ through the post-selection association factor $a'_{ij}$. The term $E_{holo}$ is the finite-change residual; it captures the non-additive information arising from the simultaneous presence of marginal and joint-structure changes. Note that $A_i$ and $B_j$ are weighted averages of the association factors over the possible partners; they are the marginal effects of host type $i$ and microbial type $j$. Thus $J_H$ and $J_M$ measure the change in the marginal representation of hosts and microbes. Furthermore, only the products $W_{ij}$ are identifiable from frequency data: for any positive constants $c_i$ and $d_j$, the factors

$$W_H^*(i)=c_iW_H(i),\quad W_M^*(j)=d_jW_M(j),\quad a_{ij}^*=\frac{a_{ij}}{c_id_j}$$

yield the same $W_{ij}$. The quantities entering the partition are, however, invariant under this reparametrization,

$$A_i^*=\frac{A_i}{c_i},\quad B_j^*=\frac{B_j}{d_j},\quad a'^*_{ij}=\frac{a_{ij}^*\bar{W}}{A_i^*B_j^*}=a'_{ij}=\frac{q'_{ij}}{p'_im'_j}$$

so $J_H$, $J_M$, $J_{assoc}$, $E$ and the tests below are well defined, and $a'_{ij}$ can be computed directly from the observed frequencies as $q'_{ij}$ / ($p'_im'_j$). Moreover, $a'_{ij}$ = 1 for all $i$, $j$ if and only if log $W_{ij}$ is additive in a host term and a microbial term, that is, if there is no host-microbiome interaction in fitness.

Consequently, different decompositions of $W_{ij}$ into $W_H$, $W_M$ and $a_{ij}$ cannot be distinguished from frequency data, because they yield the same partition. As in Carvajal-Rodríguez (2020), where one reference parameter is fixed to 1 and each candidate model assigns parameters to sets of cells, the mechanistic parameters can be estimated only within a parametric family that removes this freedom, and models that generate the same $W_{ij}$ remain indistinguishable from frequency data alone.

### *2.2.2 Statistical Tests for Selection in the Aggregate-Microbiome Model*

For a total of $n$ host-microbiome pairings, the statistic $nJ_T$ asymptotically follows a chi-squared distribution with ($K_HK_M - 1$) degrees of freedom. The divergence components of the partition can be tested analogously: $nJ_H$ follows a chi-squared distribution with $K_H - 1$ degrees of freedom, $nJ_M$ with $K_M - 1$ degrees of freedom, and $nJ_{assoc}$ with ($K_H - 1$)($K_M - 1$) degrees of freedom. These degrees of freedom sum to those of the total statistic. The partition of a total divergence into marginal and association components has classical antecedents in the partition of chi-square in contingency tables (Lancaster, 1951) and in the decomposition of information in multivariate tables (Kullback, 1997; McGill, 1954; Watanabe, 1960); TVI differs in that the reference is a random-assembly distribution, the divergence is the mean change of the character log $W$, and the components are interpreted as marginal change and association generated by the relative-change function $W$ (selection when $W$ represents fitness). The residual term $E_{\text{holo}}$ is not an independent divergence component and is therefore not assigned separate degrees of freedom. It measures the finite-change coupling between marginal and associative change: $r_{ij} - q_{ij}$ represents the change attributable to the marginal distributions, whereas log($q'_{ij} / r_{\text{ij}}$) represents departure from the corresponding independence distribution. Thus, $E_{\text{holo}} = 0$ when either the marginal distributions remain unchanged ($r = q$) or post-selection independence is preserved ($q' = r$). $E_{\text{holo}}$ has no definite sign, so $J_{\text{H}} + J_{\text{M}} + J_{\text{assoc}}$ is not bounded by $J_T$. If the marginal and association changes are of order δ, then $J_{\text{H}}$, $J_{\text{M}}$ and $J_{\text{assoc}}$ are of order $\delta^2$, whereas $E_{\text{holo}}$ is of order $\delta^3$, because the weighted marginal means of log $a'_{ij}$ are of order $\delta^2$. Hence $n{\cdot}E_{\text{holo}}$ is asymptotically negligible. Moreover, $E_{\text{holo}}$ is not a divergence, and the degrees of freedom of $J_T$, $K_HK_M - 1$, are exhausted by those of the three divergence terms, ($K_H - 1$) + ($K_M - 1$) + ($K_H - 1$)($K_M$

− 1). No test is therefore associated with $E_{\text{holo}}$.

## 2.3 Informational Model of Selection in the Holobiont: A Structured Multicomponent Microbiome

A structured multicomponent extension of the holobiont model was introduced in the Mathematical Appendix of Carvajal-Rodríguez (2026). Here, this formulation is revised and clarified in the main text, generalized to allow association effects at the level of the complete host-microbiome configuration, and extended through the development of statistical tests and a worked example illustrating their application. In contrast to the aggregate treatment adopted in the previous section, we now represent the microbiome as a structured assemblage of $K$ microbial components, each with its own set of possible states and its own contribution to holobiont dynamics. These components may represent, for example, microbial phyla, functional guilds, or ASV-defined sublineages. A complete holobiont configuration is therefore indexed by a host type $i$ and a vector of microbial states $(j_1,\ldots,j_K)$, where $j_k$ denotes the state of the $k$-th microbiome component $M_k$.

### *2.3.1 Model Formulation and Informational Partition*

Let $p_i$ be the pre-selection frequency of host type $i$, and let $m_{k,j_k}$ be the pre-selection frequency of state $j_k$ in microbial component $M_k$. The expected frequency of the complete holobiont configuration under independent association is

$$q_{i,j_1,\ldots,j_K} = p_i \prod_{k=1}^{K} m_{k,j_k}.$$

The formulation assumes that a state is defined for each microbial component in every holobiont. When biologically meaningful, absence of a component may be represented as an additional state. By contrast, missing observations should not be treated as biological absence and require either restriction to complete observations or an appropriate missing-data treatment.

The association contribution to holobiont fitness may depend on the complete host-microbiome configuration. We therefore write

$$W_{i,\boldsymbol{j}} = W_H(i) \prod_{k=1}^{K} W_{M_k}(j_k)\, a_{i,\boldsymbol{j}}$$

where

$W_H(i)$ is the host contribution,

$W_{M_k}(j_k)$ is the contribution of state $j_k$ of microbiome component $k$ and

$a_{i,\boldsymbol{j}}$ represents the association contribution of the complete configuration formed by host type $i$ and microbial states $\boldsymbol{j} = (j_1, \ldots, j_K)$.

The pairwise-separable formulation considered in Carvajal-Rodríguez (2026) is recovered when

$$a_{i,\boldsymbol{j}} = \prod_{k=1}^{K} a_{i\,j_k}^{(k)}$$

where the superscript ($k$) indicates the microbial component to which the association factor refers and

$a_{i\,j_k}^{(k)}$ is the association factor between host type $i$ and state $j_k$ of microbial component $M_k$.

The post-selection frequencies follow the standard replicator update:

$$q'_{i,j_1,\ldots,j_K} = q_{i,j_1,\ldots,j_K} \cdot \frac{W_{i,j_1,\ldots,j_K}}{\bar{W}},$$

where $\bar{W}$ is the mean fitness of the holobiont population.

Therefore, the relative fitness of each holobiont configuration can equivalently be expressed as the ratio of its post- and pre-selection frequencies:

$$\frac{W_{i,j_1,\ldots,j_K}}{\bar{W}} = \frac{q'_{i,j_1,\ldots,j_K}}{q_{i,j_1,\ldots,j_K}}.$$

As before, the character $Z$ associated with a holobiont configuration is defined as the logarithm of its relative fitness. Hence,

$$Z_{i,j_1,\ldots,j_K} = \log \frac{W_{i,j_1,\ldots,j_K}}{\bar{W}} = \log \frac{q'_{i,j_1,\ldots,j_K}}{q_{i,j_1,\ldots,j_K}}.$$

Consequently, the average change in $Z$ caused by selection is the Jeffreys divergence between the pre- and post-selection frequency distributions:

$$J_T = \sum_{i,j_1,\dots,j_K} \left( q'_{i,j_1,\dots,j_K} - q_{i,j_1,\dots,j_K} \right) \log \frac{q'_{i,j_1,\dots,j_K}}{q_{i,j_1,\dots,j_K}} \qquad (1).$$

The general fitness model gives the additive representation

$$Z_{i,\boldsymbol{j}} = \log W_H(i) + \sum_{k=1}^{K} \log W_{M_k}(j_k) + \log a_{i,\boldsymbol{j}} - \log \bar{W}$$

Let the post-selection host marginal frequency be

$$p'_i = \sum_{j_1,\dots,j_K} q'_{i,j_1,\dots,j_K},$$

and for each fixed state $j_k$ of microbial component $M_k$, its post-selection marginal frequency is obtained by summing $q'_{i,j1},\dots,j_K$ over all host types $i$ and over every combination of states of the remaining $K - 1$ microbial components. The index $j_k$ is held fixed and is therefore not included among the summation indices:

$$m'_{k,j_k} = \sum_{i=1}^{K_H} \sum_{j_1=1}^{K_{M_1}} \cdots \sum_{j_{k-1}=1}^{K_{M_{k-1}}} \sum_{j_{k+1}=1}^{K_{M_{k+1}}} \cdots \sum_{j_K=1}^{K_{M_K}} q'_{i,j_1,\dots,j_K}.$$

The corresponding pre-selection marginals satisfy

$$p_i = \sum_{j_1,\dots,j_K} q_{i,j_1,\dots,j_K},$$

Similarly, for each fixed state $j_k$ of microbial component $M_k$, its pre-selection marginal frequency is obtained by summing $q_{i,j1,\dots,jK}$ over all host types $i$ and over every combination of states of the remaining $K - 1$ microbial components. Thus, $j_k$ is held fixed and is not included among the summation indices:

$$m_{k,j_k} = \sum_{i=1}^{K_H} \sum_{j_1=1}^{K_{M_1}} \cdots \sum_{j_{k-1}=1}^{K_{M_{k-1}}} \sum_{j_{k+1}=1}^{K_{M_{k+1}}} \cdots \sum_{j_K=1}^{K_{M_K}} q_{i,j_1,\dots,j_K}$$

We then define the independent distribution having the same marginal frequencies as the post-

selection distribution:

$$r_{i,j_1,\dots,j_K} = p'_i \prod_{k=1}^{K} m'_{k,j_k}.$$

The total frequency change and the corresponding logarithmic ratio can be written as

$$q' - q = (q' - r) + (r - q)$$

and

$$\log \frac{q'}{q} = \log \frac{q'}{r} + \log \frac{r}{q}.$$

Substitution into the expression for $J_T$ produces four terms:

$$J_T = \sum (r - q) \log \frac{r}{q} + \sum (r - q) \log \frac{q'}{r} + \sum (q' - r) \log \frac{r}{q} + \sum (q' - r) \log \frac{q'}{r}.$$

The third term is exactly zero:

$$\sum (q' - r) \log \frac{r}{q} = 0.$$

Indeed,

$$\log \frac{r_{i,j_1,\dots,j_K}}{q_{i,j_1,\dots,j_K}} = \log \frac{p'_i}{p_i} + \sum_{k=1}^{K} \log \frac{m'_{k,j_k}}{m_{k,j_k}},$$

and $q'$ and $r$ have identical host and microbial-component marginal distributions. The summation of ($q'$ − $r$) against each marginal log-ratio therefore vanishes.

The first term separates exactly into the host and microbial marginal divergences:

$$\sum (r - q) \log \frac{r}{q} = J_H + \sum_{k=1}^{K} J_{M_k},$$

where

$$J_H = \sum_i (p'_i - p_i) \log \frac{p'_i}{p_i},$$

and

$$J_{M_k} = \sum_{j_k} \left( m'_{k,j_k} - m_{k,j_k} \right) \log \frac{m'_{k,j_k}}{m_{k,j_k}}.$$

The joint-structure component is

$$J_{assoc} = \sum_{i,j_1,\ldots,j_K} \left( q'_{i,j_1,\ldots,j_K} - r_{i,j_1,\ldots,j_K} \right) \log \frac{q'_{i,j_1,\ldots,j_K}}{r_{i,j_1,\ldots,j_K}}.$$

For the pairwise-separable fitness model used in Carvajal-Rodríguez (2026),

$$a_{i,\boldsymbol{j}} = \prod_{k=1}^{K} a_{i\,j_k}^{(k)}, \quad q'_{i,j_k} = \sum_{\boldsymbol{j}_{-k}} q'_{i,\boldsymbol{j}}$$

where the superscript ($k$) indicates the microbial component to which the association factor refers, and $q'_{i,j_k}$ is obtained by summing $q'_{i,j}$ over the states of the other components. The microbial components are then conditionally independent given the host type after selection, because $q'_{ij}$ factorizes, for fixed $i$, into terms that each depend on a single $j_k$. In this case the global association term decomposes exactly into the sum of the host-component association divergences,

$$J_{\text{assoc}} = \sum_{k=1}^{K} J_{\text{assoc},k}$$

where

$$J_{\text{assoc},k} = \sum_{i,j_k} \left( q'_{i,j_k} - p'_i m'_{k,j_k} \right) \log \frac{q'_{i,j_k}}{p'_i m'_{k,j_k}}$$

In its general form, $J_{\text{assoc}}$ does not attempt to distinguish associations between the host and particular microbial components from dependencies among microbial components. Rather, it quantifies the information contained in the joint structure of complete holobiont configurations beyond that explained by the marginal distributions of the host and each microbial component. It therefore generalizes the association, or “choice,” component of the two-dimensional model to a structured multicomponent microbiome. Without separability, the sum of the host-component terms need not exhaust $J_{\text{assoc}}$. Their difference defines a signed remainder, $R_{\text{cond}}$, arising from conditional dependence among microbial components given the host type. A non-zero $R_{\text{cond}}$ implies conditional dependence, but $R_{\text{cond}} = 0$ does not necessarily imply conditional independence. As a signed residual rather than a divergence, $R_{\text{cond}}$ is not assigned a separate chi-squared test. Conditional independence is instead tested directly using $G_{\text{cond}}$, introduced in Section 2.3.2 and developed in the Supplementary Mathematical Appendix.

Finally, the residual term is

$$E_{multi} = \sum_{i, j_1, \dots, j_K} \left( r_{i, j_1, \dots, j_K} - q_{i, j_1, \dots, j_K} \right) \log \frac{q'_{i, j_1, \dots, j_K}}{r_{i, j_1, \dots, j_K}} .$$

The complete informational partition is therefore

$$J_T = \left( J_H + \sum_{k=1}^{K} J_{M_k} \right) + J_{assoc} + E_{multi} \qquad (2).$$

Thus, the information generated by holobiont selection separates into a host marginal component, a set of microbial-component marginal contributions, a component associated with the joint structure of complete holobiont configurations, and a residual coupling term. The term $E_{\text{multi}}$ captures the non-additive information arising from the simultaneous presence of marginal and joint-structure changes. It is the multicomponent extension of the holobiont residual $E_{\text{holo}}$ in the aggregate-microbiome model.

Equation (2) reduces to the aggregate-microbiome holobiont model when $K = 1$, and recovers the informational partition for non-random mating (Carvajal-Rodríguez, 2018) when the two interacting components correspond to the sexes rather than to the host and microbiome.

### *2.3.2 Statistical Tests for Selection in the Multicomponent-Microbiome Model*

For a total of $n$ complete holobiont configurations, let $K_H$ denote the number of host types and $K_{M_k}$ the number of states of microbial component $M_k$, and $K$ is the total number of microbial components. The total number of possible microbiome configurations is

$$K_M = \prod_{k=1}^{K} K_{M_k} .$$

Accordingly, under the null model, the statistic $nJ_T$ asymptotically follows a chi-squared distribution

$$n\,J_T \approx \chi^2_{\nu_J},$$

with

$$\nu_J = K_H K_M - 1$$

degrees of freedom.

The marginal divergence components can be tested analogously: $nJ_H$ follows a chi-squared distribution with $K_H - 1$ degrees of freedom, whereas each $nJ_{M_k}$ follows a chi-squared distribution with $K_{M_k} - 1$ degrees of freedom.

$$nJ_H \approx \chi^2_{K_H-1};\, nJ_{M_k} \approx \chi^2_{K_{M_k}-1}.$$

The statistic $nJ_{assoc}$, which measures the joint structure of complete holobiont configurations beyond that explained by the marginal distributions of the host and the individual microbial components, follows a chi-squared distribution

$$n\,J_{assoc} \approx \chi^2_{\nu_{assoc}},$$

with

$$\nu_{assoc} = K_H \prod_{k=1}^{K} K_{M_k} - 1 - \left[ (K_H - 1) + \sum_{k=1}^{K} (K_{M_k} - 1) \right] = K_H (K_M - 1) - \sum_{k=1}^{K} (K_{M_k} - 1)$$

degrees of freedom.

Under the pairwise-separable fitness model $\left(a_{i,\mathbf{j}} = \prod_{k=1}^{K} a^{(k)}_{i\,j_k}\right)$, if the global association component is significant, the previously defined host-component association terms $J_{assoc,k}$ can be used in a post hoc analysis to identify which microbial components show association with host type. In this model, these terms provide an exact decomposition of the global association term. For each component $M_k$, the statistic $nJ_{assoc,k}$ asymptotically follows a chi-squared distribution with $(K_H - 1)(K_{M_k} - 1)$ degrees of freedom. Because several component-specific tests may be performed, adjustment for multiple comparisons should be considered.

Under the general non-separable model, the same pairwise tests can identify host-component associations, but they need not exhaust the global association information. Conditional dependence among microbial components given the host type can also be tested directly using the likelihood-ratio statistic

$$G_{\text{cond}} = 2\sum_{i=1}^{K_H} n_i \sum_j q'(j|i)\log\left(\frac{q'(j|i)}{\prod_{k=1}^{K} q'(j_k|i)}\right)$$

where $n_i$ is the observed number of holobionts of host type $i$, and $q'(j \mid i)$ and $q'(j_k \mid i)$ denote the joint and component-specific microbial distributions conditional on host type $i$, respectively. Under the null hypothesis of conditional independence, $G_{\text{cond}}$ asymptotically follows a chi-squared distribution,

$$G_{\text{cond}} \approx \chi^2_{\nu_{\text{cond}}}$$

where

$$\nu_{\text{cond}} = K_H\left[K_M - 1 - \sum_{k=1}^{K}(K_{M_k} - 1)\right]$$

The degrees of freedom of the global association test are partitioned exactly into those of the host-component association tests and the conditional-independence test:

$$\nu_{\text{assoc}} = \sum_{k=1}^{K} \nu_{\text{assoc},k} + \nu_{\text{cond}}$$

The host-component association tests provide an exact decomposition of $J_{\text{assoc}}$ only under the pairwise-separable model, whereas $G_{\text{cond}}$ tests conditional dependence among microbial components within host types. As a likelihood-ratio statistic, $G_{\text{cond}}$ is distinct from $R_{\text{cond}}$ and is not generally equal to $nR_{\text{cond}}$. Its derivation and illustrative applications are provided in the Supplementary Mathematical Appendix.

The null hypothesis depends on the test considered. For $nJ_T$, it is that the relative-change function is identically one, whereas the marginal tests concern unchanged host or microbial-component frequencies. The association tests evaluate post-selection independence among all components ($nJ_{\text{assoc}}$), between the host and individual microbial components ($nJ_{\text{assoc,k}}$), or among microbial components conditional on host type ($G_{\text{cond}}$). The biological interpretation of a rejection depends on how $W$ is specified: if $W$ represents fitness, the change is attributed to selection; if the model incorporates other processes in $W$ (migration, transmission), the change is attributed to them. Attributing the change to selection therefore requires that $W$ be fitness, either by design (for example, an experiment in which all host-microbiome combinations are assembled in known proportions, such as germ-free hosts inoculated with defined microbiomes, and then exposed to the same environment) or by specification. The tests assume a deterministic null plus multinomial sampling error: if there is drift between the reference and the observed distributions, the real variance exceeds the nominal one and the tests will tend to reject more often than expected. Averaging across independent replicates may reduce stochastic fluctuations, but formal calibration of the tests under drift requires an appropriate stochastic null model, which is not developed here. Finally, the divergence estimated from a sample is biased upwards by approximately $v / n$ under the null (the mean of the chi-squared distribution divided by $n$), so the estimated $J$ should not be read as an effect size without correction; the $p$-values are not affected because the reference distribution already includes this bias.

When the reference marginal frequencies are estimated from a reference sample of size $n_0$ rather than known, the variance of the difference between observed and reference marginals is larger than that obtained treating them as fixed. For the marginal components $J_H$ and $J_{M_k}$, this is approximately accounted for by replacing $n$ with an effective sample size, $n_e = nn_0 / (n + n_0)$. This is an approximation; when $J_T$ is computed with estimated marginals, calibration by Monte Carlo is safer. $J_{\text{assoc}}$ is not affected, because it uses the post-selection marginals of the sample itself.

As in the aggregate model, $E_{\text{multi}}$ is a finite-change coupling term rather than an independent divergence component and therefore carries no additional degrees of freedom. For empirical applications, these asymptotic chi-square approximations require adequate sampling across the relevant configuration space. Structural zeros, corresponding to biologically impossible configurations, should be excluded from the set of possible configurations, with the corresponding adjustment of the degrees of freedom, whereas sparse sampling may require pooling biologically meaningful classes, where appropriate, or calibrating the null distribution by Monte Carlo simulation.

## 2.4 A Toy Example Illustrating the Multicomponent-Microbiome Statistical Tests

Consider a sample of $n$ = 1080 octopus holobionts comprising three host size types: small (S), medium (M), and large (L), and two microbial components, $M_1$ and $M_2$. For illustrative purposes, $M_1$ may be taken to represent the skin-associated microbial community and $M_2$ the gut-associated microbial community. Each component has two alternative states or profiles, denoted A and B. The four possible microbial configurations are ($M_{1,\mathrm{A}}M_{2,\mathrm{A}}$), ($M_{1,\mathrm{A}}M_{2,\mathrm{B}}$), ($M_{1,\mathrm{B}}M_{2,\mathrm{A}}$), ($M_{1,\mathrm{B}}M_{2,\mathrm{B}}$).

Because $K_H = 3$, $K_{M1} = K_{M2} = 2$, there are $K_H K_{M1} K_{M2} = 3 \times 2 \times 2 = 12$ possible complete holobiont configurations. For example, ($H_\mathrm{S}M_{1,\mathrm{A}}M_{2,\mathrm{B}}$) represents a small host associated with state A of microbial component $M_1$ and state B of microbial component $M_2$.

Assume that the frequencies of the complete host-microbiome configurations have been recorded in the focal sample. Independently, the marginal reference frequencies of the host types and microbial-component states are known or estimated from an appropriate reference population and treated as fixed for the purposes of the present test. Depending on the system, these may include a preceding or parental host population, gametic or genetic-marker information, microbial source pools, or other suitable reference samples. These marginal frequencies define the distribution of complete configurations expected under random association. In the present example, the three host types occur at equal marginal frequencies:

$$p_S = p_M = p_L = \frac{1}{3}.$$

Likewise, the two states of each microbial component are assumed to have been equally represented in the corresponding microbial source pools available during holobiont formation:

$$m_{1,A} = m_{1,B} = m_{2,A} = m_{2,B} = \frac{1}{2}.$$

If complete holobiont configurations are formed independently according only to these marginal frequencies, their expected frequencies are

$$q_{i,j_1,j_2} = p_i m_{1,j_1} m_{2,j_2} = \frac{1}{3} \cdot \frac{1}{2} \cdot \frac{1}{2} = \frac{1}{12}, \text{ for every } (i \in \mathrm{S,M,L}) \wedge (j_1, j_2 \in \mathrm{A,B}).$$

Thus, under the random-association reference, each of the 12 possible holobiont configurations has an expected count of

$$n q_{i,j_1,j_2} = 1080 \cdot \frac{1}{12} = 90.$$

These are expected counts under independent holobiont assembly, rather than observed counts. Differential success may arise during holobiont assembly, when particular host-microbiome combinations establish more successfully than expected from their marginal frequencies, or during subsequent survival. After the processes determining the differential realization of holobiont configurations, their frequencies are recorded to obtain *q′* (Table 1). These observed frequencies may therefore reflect differences in the relative success of host types, microbial-component states, or complete holobiont configurations, including preferential associations among particular host-microbiome combinations. The observed distribution *q′* is then compared with the random-association reference distribution *q* to determine which informational components account for the observed departure.

**Table 1. Observed counts of octopus holobiont configurations by host size type and skin- and gut-associated microbial profiles.**

| Host type | $M_{1,A}M_{2,A}$ | $M_{1,A}M_{2,B}$ | $M_{1,B}M_{2,A}$ | $M_{1,B}M_{2,B}$ | Total |
|---|---|---|---|---|---|
| Small | 160 | 80 | 80 | 40 | 360 |
| Medium | 90 | 90 | 90 | 90 | 360 |
| Large | 40 | 80 | 80 | 160 | 360 |
| Total | 290 | 250 | 250 | 290 | 1080 |

The observed post-selection frequencies are therefore obtained as

$$q'_{i,j_1,j_2} = \frac{N'_{i,j_1,j_2}}{n},$$

where $N'_{i,j_1,j_2}$ is the observed count of each complete configuration. For example $q'_{S1A2A}$ = 160 / 1080 = 0.148148.

The relative fitness of each configuration, understood broadly to include its success during holobiont assembly as well as subsequent survival or reproduction, can then be estimated directly from the replicator relation:

$$\frac{\hat{W}_{i,j_1,j_2}}{\bar{W}} = \frac{q'_{i,j_1,j_2}}{q_{i,j_1,j_2}} = \frac{N'_{i,j_1,j_2}}{90}.$$

This gives the following estimated relative fitnesses (Table 2):

**Table 2. Estimated relative fitnesses of complete holobiont configurations, calculated as $q'_{i,j_1,j_2}$ / $q_{i,j_1,j_2}$.**

| Host type | $M_{1,A}M_{2,A}$ | $M_{1,A}M_{2,B}$ | $M_{1,B}M_{2,A}$ | $M_{1,B}M_{2,B}$ |
|---|---|---|---|---|
| Small | 16/9 | 8/9 | 8/9 | 4/9 |
| Medium | 1 | 1 | 1 | 1 |
| Large | 4/9 | 8/9 | 8/9 | 16/9 |

Thus, configurations containing state A of both microbial components have the highest estimated relative fitness in small hosts, whereas configurations containing state B of both components have the highest relative fitness in large hosts. Medium-sized hosts show no detectable fitness differences among microbial configurations.

Table 3 shows the contribution of each complete host-microbiome configuration to the total Jeffreys divergence. Because $q_{i,j_1,j_2}$ = 1 / 12 for every configuration, each contribution is calculated by subtracting 1 / 12 from the corresponding observed frequency $q'_{i,j_1,j_2}$ and multiplying the resulting difference by log($q'_{i,j_1,j_2}$ / $q_{i,j_1,j_2}$). Summing the entries in the final column gives the total Jeffreys divergence (1) between the observed and random association distributions.

**Table 3. Configuration-specific contributions to the Jeffreys divergence $J_T$ in the octopus example.**

| Host type | Microbial configuration | $q'_{i,j_1,j_2}$ | $q'_{i,j_1,j_2} - q_{i,j_1,j_2}$ | $q'_{i,j_1,j_2} / q_{i,j_1,j_2}$ | $(q' - q) \log(q' / q)$ |
|---|---|---|---|---|---|
| Small | $M_{1,A}M_{2,A}$ | 0.148148 | 0.064815 | 16 / 9 | 0.037292 |
| Small | $M_{1,A}M_{2,B}$ | 0.074074 | −0.009259 | 8 / 9 | 0.001091 |
| Small | $M_{1,B}M_{2,A}$ | 0.074074 | −0.009259 | 8 / 9 | 0.001091 |
| Small | $M_{1,B}M_{2,B}$ | 0.037037 | −0.046296 | 4 / 9 | 0.037543 |
| Medium | $M_{1,A}M_{2,A}$ | 0.083333 | 0 | 1 | 0 |
| Medium | $M_{1,A}M_{2,B}$ | 0.083333 | 0 | 1 | 0 |
| Medium | $M_{1,B}M_{2,A}$ | 0.083333 | 0 | 1 | 0 |
| Medium | $M_{1,B}M_{2,B}$ | 0.083333 | 0 | 1 | 0 |
| Large | $M_{1,A}M_{2,A}$ | 0.037037 | −0.046296 | 4 / 9 | 0.037543 |
| Large | $M_{1,A}M_{2,B}$ | 0.074074 | −0.009259 | 8 / 9 | 0.001091 |
| Large | $M_{1,B}M_{2,A}$ | 0.074074 | −0.009259 | 8 / 9 | 0.001091 |
| Large | $M_{1,B}M_{2,B}$ | 0.148148 | 0.064815 | 16 / 9 | 0.037292 |
| **Total** | | **1** | **0** | — | **0.154033** |

Thus

$$J_T = \sum_{i,j_1,j_2} \left(q'_{i,j_1,j_2} - q_{i,j_1,j_2}\right) \log \frac{q'_{i,j_1,j_2}}{q_{i,j_1,j_2}} = 0.154033 .$$

The corresponding global statistic is

$$n\,J_T = 1080 \times 0.154033 = 166.355 .$$

The total number of degrees of freedom is

$$\nu_J = K_H K_{M_1} K_{M_2} - 1 = 3 \cdot 2 \cdot 2 - 1 = 11.$$

Thus,

$$n\,J_T \approx \chi^2_{11} ,$$

giving

$$P = 6.63 \times 10^{-30} .$$

The null hypothesis that the observed distribution of complete holobiont configurations does not differ from that expected under random association is therefore strongly rejected.

To determine whether this departure from random association arises from changes in the host or microbial-component marginal distributions, from a reorganization of their joint structure, or from both, we next examine the informational partition of $J_T$ (2)

Each host type is represented by 360 holobionts in the observed population (Table 1), so that

$$p'_i = p_i = \frac{1}{3}.$$

Similarly, each state of each microbial component occurs 540 times:

$$m'_{1,A} = m'_{1,B} = m'_{2,A} = m'_{2,B} = \frac{1}{2}.$$

Consequently, the post-selection independence reference is identical to the pre-selection random-association distribution:

$$r_{i,j_1,j_2} = p'_i m'_{1,j_1} m'_{2,j_2} = q_{i,j_1,j_2}.$$

Therefore, $J_H = 0$, $J_{M1} = J_{M2} = 0$. Accordingly, the corresponding marginal statistics are zero and non-significant (Table 4).

Since $r = q$, the association divergence is equal to the total divergence:

$$J_{assoc} = \sum_{i,j_1,j_2} \left(q'_{i,j_1,j_2} - r_{i,j_1,j_2}\right) \log \frac{q'_{i,j_1,j_2}}{r_{i,j_1,j_2}} = 0.154033.$$

The association degrees of freedom are

$$\nu_{assoc} = K_H K_{M_1} K_{M_2} - 1 - \left[\left(K_H - 1\right) + \left(K_{M_1} - 1\right) + \left(K_{M_2} - 1\right)\right],$$

and hence

$$\nu_{assoc} = 11 - \left(2 + 1 + 1\right) = 7.$$

The corresponding statistic is therefore

$$n J_{assoc} = 166.355 \approx \chi^2_7,$$

with

$$P = 1.47 \cdot 10^{-32}.$$

The joint-structure component is thus highly significant. The selection response is not caused by changes in the overall frequencies of host size classes or microbial states, but by a reorganization of the complete holobiont configurations: small hosts become preferentially associated with the $M_{1,A}M_{2,A}$ configuration, whereas large hosts become preferentially associated with $M_{1,B}M_{2,B}$. Host size is treated here as a host type. If size were itself influenced by the microbiome during development, the observed association would not imply selection on the combination. Finally, because the pre- and post-selection marginal distributions coincide, ($r = q$), it follows that $E_{multi} = 0$. The informational partition therefore reduces in this example to $J_T = J_{\text{assoc}}$.

**Table 4. Total and partitioned informational change and corresponding chi-square tests in the octopus holobiont example.**

| Informational term | Interpretation | Divergence | Test statistic | Degrees of freedom | P-value |
|---|---|---|---|---|---|
| $J_T$ | Total change in complete holobiont configurations | 0.154033 | $nJ_T = 166.355$ | $K_H K_{M1} K_{M2} - 1 = 11$ | $6.63\times10^{-30}$ |
| $J_H$ | Host marginal component | 0 | $nJ_H = 0$ | $K_H - 1 = 2$ | 1 |
| $J_{M1}$ | Marginal component of $M_1$ | 0 | $nJ_{M1} = 0$ | $K_{M1} - 1 = 1$ | 1 |
| $J_{M2}$ | Marginal component of $M_2$ | 0 | $nJ_{M2} = 0$ | $K_{M2} - 1 = 1$ | 1 |
| $J_{assoc}$ | Joint-structure association component | 0.154033 | $nJ_{\text{assoc}} = 166.355$ | $11 - (2+1+1) = 7$ | $1.47\times10^{-32}$ |
| $E_{multi}$ | Residual non-additive component | 0 | — | — | — |

This example illustrates how observed configuration counts can be used to test and partition selection-associated informational change. Here, the effect is entirely attributable to joint holobiont structure, whereas in general the same framework can separately detect changes in host and microbial-component marginals and in their association. Because the fitness table of this example is a product over components, the pairwise-separable model holds exactly and the host-component association terms add up to the global association term; they are reported in the Supplementary Mathematical Appendix, together with a second example that is not pairwise-separable and in which the remainders $E_{\text{multi}}$ and $R_{\text{cond}}$ are non-zero.

## 3 Multicomponent Sets Under Within- and Between-Group Selection

The preceding holobiont model can be extended to a population of multicomponent groups. At this level, each observational unit consists of a single group, which belongs to one of $G$ mutually exclusive group classes, indexed by $g = 1,..., G$, with frequencies $p_g$. Thus, although the population may contain many groups belonging to different classes, each individual observation is characterized by a single group-class index $g$.

A group of class $g$ contains $K_g$ components, indexed by $k = 1, \dots, K_g$. Component $k$ can occur in $K_{g,k}$ possible states, indexed by $j_{g,k} = 1, \dots, K_{g,k}$. A complete internal configuration of a group of class $g$ is therefore represented by the vector

$$\boldsymbol{j}_g = \left(j_{g,1}, \dots, j_{g,K_g}\right),$$

with frequency $q_{g,\boldsymbol{j}_g}$ among groups of that class.

Selection may act simultaneously at two hierarchical levels: among group classes and among internal configurations within each group class. Accordingly, using $\omega$ to denote relative fitness (so that, for a complete state, $\omega = \frac{W}{\bar{W}}$ ), the relative fitness of a complete state can be factorized as

$$\omega_{g,\boldsymbol{j}_g} = \omega_{I,g,\boldsymbol{j}_g} \cdot \omega_{G,g},$$

where $\omega_{I,g,\boldsymbol{j}_g}$ denotes the relative fitness of internal configuration $\boldsymbol{j}_g$ within group class $g$, and $\omega_{G,g}$ denotes the relative collective fitness of group class $g$. Equivalently, $\omega_{G,g}$ is the mean relative fitness of group class $g$, obtained by averaging over its internal configurations according to their frequencies. The two factors are normalized at their respective levels:

$$\sum_{\boldsymbol{j}_q} q_{g,\boldsymbol{j}_g} \omega_{I,g,\boldsymbol{j}_g} = 1$$

for every $g$, and

$$\sum_{g=1}^{G} p_g \omega_{G,g} = 1.$$

The group frequencies and the within-group-class frequencies of their internal configurations then change according to

$$p'_g = p_g \omega_{G,g}$$

and

$$q'_{g,\boldsymbol{j}_g} = q_{g,\boldsymbol{j}_g}\,\omega_{I,g,\boldsymbol{j}_g},$$

respectively.

The informational change generated between group classes is therefore

$$J_G = \sum_{g=1}^{G} \left(p'_g - p_g\right) \log \frac{p'_g}{p_g},$$

whereas the informational change within group class $g$ is

$$J_{I,g} = \sum_{\boldsymbol{j}_g} \left(q'_{g,\boldsymbol{j}_g} - q_{g,\boldsymbol{j}_g}\right) \log \frac{q'_{g,\boldsymbol{j}_g}}{q_{g,\boldsymbol{j}_g}}.$$

Under the same initial-independence assumption adopted above, the multicomponent internal distribution admits the same type of partition developed for the holobiont:

$$J_{I,g} = \left(\sum_{k=1}^{K_g} J_{g,k}\right) + J_{g,\text{assoc}} + E_g,$$

where the first term measures informational changes in the marginal distributions of the components, $J_{g,\text{assoc}}$ is the Jeffreys divergence associated with post-selection departures from independence among components, and $E_g$ is the finite-change residual. The explicit definitions and derivation of this partition are given in the Supplementary Mathematical Appendix.

The total informational change can also be evaluated from the joint distribution of group classes and their internal configurations before and after selection. Defining

$$F_{g,\boldsymbol{j}_g} = p_g \cdot q_{g,\boldsymbol{j}_g}$$

the corresponding Jeffreys divergence can be written exactly as

$$J_T = J_G + \sum_{g=1}^{G} \bar{p}_g \cdot J_{I,g} + E_{GI}$$

where

$$\bar{p}_g = \frac{p'_g + p_g}{2}$$

and $E_{GI}$ is a hierarchical coupling term arising because the forward and reverse components of internal informational change are weighted by the post- and pre-selection group-class frequencies, respectively. Importantly, $J_{I,g}$ remains the Jeffreys divergence describing local informational change within group class $g$. The weighting appearing in the total partition does not modify this local interpretation. It arises because the internal distribution is conditional on group class $g$, whose frequency also changes. Consequently, the forward internal Kullback-Leibler contribution is weighted by the post-selection frequency $p'_g$, whereas the reverse contribution is weighted by the pre-selection frequency $p_g$. Their symmetric representation yields the mean-weighted local Jeffreys term together with the hierarchical coupling term $E_{GI}$. Its explicit form and the conditions under which it vanishes or becomes negligible are derived in the Supplementary Mathematical Appendix.

This formulation contains as special cases the informational partition developed by Carvajal-Rodríguez (2026) for the tragedy-of-the-commons model presented by Frank (2025), and the aggregate and multicomponent holobiont models developed above. More importantly, it reveals a recursive structure: units at one organizational level may themselves consist of multicomponent units whose marginal distributions and associations can undergo further informational change. The next section generalizes this structure to an arbitrary number of nested organizational levels.

## 4 Hierarchical Multicomponent Sets

The preceding model can be viewed as the case in which each higher-level unit contains a single multicomponent group. Group classes are mutually exclusive alternatives for that group, so their frequencies can change under selection, but different group classes do not jointly occur within the same observational unit and their association cannot be explicitly represented.

The hierarchical extension removes this restriction by allowing each higher-level set to contain $L \geq 1$ constituent groups. Each position $l = 1,..., L$ is occupied by a group belonging to one of the $G$ possible group classes, with its class denoted by $g_l$. Consequently, the vector $\boldsymbol{g} = (g_1,\ldots,g_L)$ specifies the joint group-class profile of the higher-level set. For example, if $G = 4$ and $L = 2$, possible profiles include (1, 1), (1, 2), (1, 3), (1, 4), etc. Because the $L$ constituent groups now occur jointly within the same higher-level unit, associations among their group classes can be explicitly represented.

Associations may then change at each level of the hierarchy: among components within groups, as in holobionts; among the internal states of different constituent groups; and among the group classes that co-occur within higher-level sets. These nested forms of association provide the formal basis for the Theory of Variable Interactions.

## 4.1 Hierarchical States, Classes, and Frequencies

Consider a population of higher-level sets, each composed of $L$ constituent groups occupying distinguishable positions, roles, or functional units, indexed by $l = 1,..., L$. The index $l$ identifies the position of a constituent group within the higher-level set; it does not specify the class to which that group belongs. Each constituent group belongs to one of $G$ possible group classes. The class of the group occupying position $l$ is denoted by $g_l$, with $g_l = 1, \dots, G$. Different positions may be occupied by groups belonging to the same class; thus, the $L$ constituent groups need not belong to $L$ different classes. For simplicity of notation, the same number $G$ of possible group classes is assumed at all positions; this assumption is not essential to the formulation.

The vector $\boldsymbol{g} = (g_1,..., g_L)$ specifies the ordered group-class profile of the higher-level set. For example, $\boldsymbol{g} = (2, 1, 2)$ describes a set in which the groups occupying positions 1 and 3 both belong to group class 2, whereas the group occupying position 2 belongs to group class 1. Importantly, membership in the same group class does not imply identical internal configurations. A group class represents a broader category, such as a species, genotype, ecological type, or holobiont type, whereas different groups belonging to that class may display different states of their internal components. Thus, the class-2 groups occupying positions 1 and 3 in the example above may have different internal configurations.

Higher-level sets sharing the same vector $\boldsymbol{g}$ therefore have the same group-class profile, although they may differ in the internal configurations of one or more of their constituent groups. Let $u_{\boldsymbol{g}}$ denote the frequency of higher-level sets with group-class profile $\boldsymbol{g}$, so that

$$\sum_{\boldsymbol{g}} u_{\boldsymbol{g}} = 1.$$

A group of class $g_l$ contains $K_{g_l}$ internal components. Component $k$ may occur in one of $K_{g_l,k}$

possible states. The internal configuration of the group of class $g_l$ occupying position $l$ is represented by

$$\boldsymbol{j}_{l,g_l}=\left(j_{l,g_l,1},\dots,j_{l,g_l,K_{g_l}}\right).$$

where

$$j_{l,g_l,k}\in\left\{1,\dots,K_{g_l,k}\right\}$$

denotes the state of internal component $k$ within the group of class $g_l$ occupying position $l$ with $j_{l,g_l,k}$ belonging to the set of possible states (from 1 to $K_{g_l,k}$). Groups belonging to the same group class therefore share the same component structure, but may differ in the states displayed by those components.

The combined internal configuration of all $L$ constituent groups is

$$\boldsymbol{j}_{\boldsymbol{g}}=\left(\boldsymbol{j}_{1,g_1},\dots,\boldsymbol{j}_{L,g_L}\right).$$

Written explicitly, this configuration is

$$\boldsymbol{j}_{\boldsymbol{g}}=\left(\left(j_{1,g_1,1},\dots,j_{1,g_1,K_{g_1}}\right),\dots,\left(j_{L,g_L,1},\dots,j_{L,g_L,K_{g_L}}\right)\right).$$

Thus, $\boldsymbol{g}$ specifies the classes of the groups occupying the $L$ positions, whereas $\boldsymbol{j}_{\boldsymbol{g}}$ specifies their particular internal configurations. The complete hierarchical state of a higher-level set is consequently represented by the pair $(\boldsymbol{g},\boldsymbol{j}_{\boldsymbol{g}})$.

Let $q_{\boldsymbol{g},\boldsymbol{j}_{\boldsymbol{g}}}$ denote the frequency of the combined internal configuration $\boldsymbol{j}_{\boldsymbol{g}}$ among higher-level sets sharing group-class profile $\boldsymbol{g}$. For every fixed group-class profile $\boldsymbol{g}$,

$$\sum_{\boldsymbol{j}_{\boldsymbol{g}}} q_{\boldsymbol{g},\boldsymbol{j}_{\boldsymbol{g}}}=1.$$

The formulation therefore distinguishes two forms of variation. Higher-level sets may differ in the classes of the constituent groups occupying their $L$ positions, as described by $\boldsymbol{g}$, and sets sharing the same group-class profile may differ in the internal states of those groups, as described by $\boldsymbol{j}_{\boldsymbol{g}}$.

The frequency in the entire population of the complete hierarchical state $(\boldsymbol{g},\boldsymbol{j}_{\boldsymbol{g}})$ is

$$F_{\boldsymbol{g},\boldsymbol{j}_{\boldsymbol{g}}}=u_{\boldsymbol{g}}\cdot q_{\boldsymbol{g},\boldsymbol{j}_{\boldsymbol{g}}}.$$

Here, $u_{\boldsymbol{g}}$ gives the frequency of the group-class profile $\boldsymbol{g}$, whereas $q_{g, j_g}$ gives the frequency of the particular combined internal configuration among sets sharing that profile. Consequently,

$$\sum_{\boldsymbol{j}_g} F_{\boldsymbol{g},\boldsymbol{j}_g} = u_{\boldsymbol{g}},$$

and

$$\sum_{\boldsymbol{g}} \sum_{\boldsymbol{j}_g} F_{\boldsymbol{g},\boldsymbol{j}_g} = 1.$$

After selection, the corresponding frequency is

$$F'_{\boldsymbol{g},\boldsymbol{j}_g} = u'_{\boldsymbol{g}} \cdot q'_{\boldsymbol{g},\boldsymbol{j}_g}.$$

## 4.2 A Simple Illustration of the Hierarchical Structure

A simple example may help clarify the different levels represented by this notation. Consider higher-level sets containing three constituent groups, $L = 3$, with two possible group classes, $G = 2$. For simplicity, suppose that both group classes contain two internal components, $K_1 = K_2 = 2$, and that each component can occur in either of two possible states. To distinguish component identity from state index in the example, we label the two states of the first component $A_1$ and $A_2$, and those of the second component $B_1$ and $B_2$.

Consider first a higher-level set with group-class profile $\boldsymbol{g} = (2, 1, 2)$. Thus, the groups occupying positions 1 and 3 belong to class 2, whereas the group occupying position 2 belongs to class 1. Suppose that their complete internal configurations are, respectively,

$$\boldsymbol{j}_{1,2} = (A_2, B_2), \boldsymbol{j}_{2,1} = (A_1, B_2), \boldsymbol{j}_{3,2} = (A_2, B_1).$$

The combined internal configuration of this higher-level set is therefore

$$\boldsymbol{j}_{\boldsymbol{g}} = ((A_2, B_2), (A_1, B_2), (A_2, B_1)).$$

Another higher-level set may share the same group-class profile $\boldsymbol{g} = (2, 1, 2)$ while displaying different internal configurations; conversely, sets may differ in their group-class profiles irrespective of their internal configurations.

These distinctions define the three nested levels of association considered below. At the uppermost level, association concerns whether particular group-class profiles occur more or less frequently than expected from the marginal frequencies of the classes occupying each position. At the intermediate level, the group-class profile is held fixed and association concerns the co-occurrence of complete internal configurations of different constituent groups. At the lowest level, association concerns the co-occurrence of component states within an individual constituent group. The following sections partition the informational effects of selection across these three levels.

## 4.3 Selection of complete hierarchical configurations

Let the normalized relative fitness of the complete hierarchical configuration be $\omega_{\boldsymbol{g},j_{\boldsymbol{g}}}$. Selection changes the frequency of each complete hierarchical configuration according to

$$F'_{\boldsymbol{g},\boldsymbol{j}_{\boldsymbol{g}}}=F_{\boldsymbol{g},\boldsymbol{j}_{\boldsymbol{g}}}\cdot\omega_{\boldsymbol{g},\boldsymbol{j}_{\boldsymbol{g}}},$$

with normalization

$$\sum_{\boldsymbol{g}}\sum_{\boldsymbol{j}_{\boldsymbol{g}}}F_{\boldsymbol{g},\boldsymbol{j}_{\boldsymbol{g}}}\cdot\omega_{\boldsymbol{g},\boldsymbol{j}_{\boldsymbol{g}}}=1.$$

The fitness of the complete hierarchical configuration can be factorized into a group-class-profile component and an internal-configuration component:

$$\omega_{\boldsymbol{g},\boldsymbol{j}_{\boldsymbol{g}}}=\omega_{G,\boldsymbol{g}}\cdot\omega_{I,\boldsymbol{g},\boldsymbol{j}_{\boldsymbol{g}}},$$

where

$$\omega_{G,\boldsymbol{g}}=\frac{u'_{\boldsymbol{g}}}{u_{\boldsymbol{g}}}$$

and

$$\omega_{I,\boldsymbol{g},\boldsymbol{j}_{\boldsymbol{g}}}=\frac{q'_{\boldsymbol{g},\boldsymbol{j}_{\boldsymbol{g}}}}{q_{\boldsymbol{g},\boldsymbol{j}_{\boldsymbol{g}}}}.$$

Thus $\omega_{G,\boldsymbol{g}}$ describes differential success among complete group-class profiles, whereas $\omega_{I,\boldsymbol{g},\boldsymbol{j}_{\boldsymbol{g}}}$ describes differential success among internal configurations within each profile.

## 4.4 Total information across the hierarchy

The total information generated by selection is the Jeffreys divergence ($J_T$) between the initial and post-selection distributions of complete hierarchical states:

$$J_T = \sum_{\boldsymbol{g}} \sum_{\boldsymbol{j}_{\boldsymbol{g}}} \left( F'_{\boldsymbol{g},\boldsymbol{j}_{\boldsymbol{g}}} - F_{\boldsymbol{g},\boldsymbol{j}_{\boldsymbol{g}}} \right) \log\left( \frac{F'_{\boldsymbol{g},\boldsymbol{j}_{\boldsymbol{g}}}}{F_{\boldsymbol{g},\boldsymbol{j}_{\boldsymbol{g}}}} \right).$$

Define the set-level information as the Jeffreys divergence ($J_S$) between the initial and post-selection distributions of higher-level sets, where sets are classified according to their group-class profiles $\boldsymbol{g}$ = ($g_1$, ..., $g_L$), as

$$J_S = \sum_{\boldsymbol{g}} \left( u'_{\boldsymbol{g}} - u_{\boldsymbol{g}} \right) \log\left( \frac{u'_{\boldsymbol{g}}}{u_{\boldsymbol{g}}} \right) \qquad (3).$$

This quantity measures the total informational change in the distribution of group-class profiles. Because each profile comprises the joint configuration of the $L$ constituent group classes, $J_S$ contains informational changes in their marginal frequencies as well as information generated by associations among them. Under the initial-independence assumption, these contributions can be partitioned, as developed below, as

$$J_S = \sum_{l=1}^{L} J_{G_l} + J_{G,\text{assoc}} + E_S,$$

where $J_{G_l}$ measures the marginal informational change in the distribution of group classes occupying position $l$ across higher-level sets, $J_{G,\text{ assoc}}$ is the Jeffreys association information generated among the constituent group classes within higher-level sets, and $E_S$ is the residual term required to complete the partition exactly.

This partition concerns changes at the level of the group-class profiles. To complete the decomposition of the total hierarchical information, it is also necessary to quantify the changes occurring in the combined internal configurations of the constituent groups within each profile. Accordingly, for each group-class profile $\boldsymbol{g}$, define the forward and reverse internal Kullback-Leibler divergences as

$$D^{+}_{I,\boldsymbol{g}}=\sum_{j_g} q'_{\boldsymbol{g},j_g}\log\left(\frac{q'_{\boldsymbol{g},j_g}}{q_{\boldsymbol{g},j_g}}\right)$$

and

$$D^{-}_{I,\boldsymbol{g}}=\sum_{j_g} q_{\boldsymbol{g},j_g}\log\left(\frac{q_{\boldsymbol{g},j_g}}{q'_{\boldsymbol{g},j_g}}\right).$$

respectively. Their sum defines the internal Jeffreys information among higher-level sets sharing the group-class profile $\boldsymbol{g}$:

$$J_{I,\boldsymbol{g}}=D^{+}_{I,\boldsymbol{g}}+D^{-}_{I,\boldsymbol{g}}.$$

The exact hierarchical decomposition of the total information is

$$J_T=J_S+\sum_{\boldsymbol{g}}\left[u'_{\boldsymbol{g}}\cdot D^{+}_{I,\boldsymbol{g}}+u_{\boldsymbol{g}}\cdot D^{-}_{I,\boldsymbol{g}}\right].$$

Equivalently, defining

$$\bar{u}_{\boldsymbol{g}}=\frac{u_{\boldsymbol{g}}+u'_{\boldsymbol{g}}}{2},$$

the total information can be written as

$$J_T=J_S+\sum_{\boldsymbol{g}}\bar{u}_{\boldsymbol{g}}\cdot J_{I,\boldsymbol{g}}+E_{SI}$$

where the hierarchical coupling term is

$$E_{SI}=\frac{1}{2}\sum_{\boldsymbol{g}}\left(u'_{\boldsymbol{g}}-u_{\boldsymbol{g}}\right)\cdot\left(D^{+}_{I,\boldsymbol{g}}-D^{-}_{I,\boldsymbol{g}}\right).$$

The appearance of these weights is a consequence of hierarchical embedding rather than of the association partition itself. For each fixed group-class profile $\boldsymbol{g}$, $J_{I,g}$ remains the local Jeffreys divergence between the pre- and post-selection internal distributions. However, when this conditional distribution is embedded in the complete hierarchical distribution, the frequency of its higher-level context also changes. The forward internal divergence is therefore weighted by $u'_{\boldsymbol{g}}$, whereas the reverse divergence is weighted by $u_{\boldsymbol{g}}$. Equivalently, their contribution can be expressed as the mean-weighted local Jeffreys divergence  plus the hierarchical coupling term $E_{SI}$. $E_{SI}$ accounts for the difference between the exact weights ($u'_g$ for the forward and $u_g$ for the reverse divergence) and the mean weight $\bar{u}_g$; it is not an additional effect of selection, but the term needed to express the total in terms of the local Jeffreys divergences.

Thus, local informational change at each hierarchical scale is represented by a Jeffreys divergence, whereas embedding lower-level change within higher-level states whose frequencies also change generates an explicit coupling between hierarchical levels. This identity does not depend on the number of levels: it applies to each level of the hierarchy, as shown in Section 5.

## 4.5 Variable associations among group classes

We first partition the set-level information $J_S$ into marginal informational changes associated with the group classes occupying each position and an association component among these constituent group classes, together with the residual required for an exact partition.

The higher-level-set profile $\boldsymbol{g} = (g_1,..., g_L)$ specifies the classes of the $L$ constituent groups. To establish the initial distribution of these profiles, let $p_{l,\,g_l}$ denote the frequency of group class $g_l$ at position $l$, with

$$\sum_{g_l=1}^{G} p_{l,g_l} = 1$$

for every position $l$. We assume that, before selection, group classes are combined independently across positions. The initial frequency of group-class profile $\boldsymbol{g}$ is therefore

$$u_{\boldsymbol{g}} = \prod_{l=1}^{L} p_{l,g_l}.$$

Thus, the initial joint distribution is completely determined by the marginal group-class frequencies and contains no association among group classes.

Selection generates the post-selection distribution $u'_g$, from which the new marginal frequencies are obtained as

$$p'_{l,g_l} = \sum_{\boldsymbol{g}_{-l}} u'_{\boldsymbol{g}},$$

where $\boldsymbol{g}_{-l}$ denotes the classes of all constituent groups except the group occupying position $l$.

The marginal distributions satisfy

$$\sum_{g_l=1}^{G} p'_{l,g_l} = 1$$

for every position *l*.

Using the post-selection marginal frequencies, define the fitted post-selection distribution under independence as

$$r'_{G,\boldsymbol{g}} = \prod_{l=1}^{L} p'_{l,g_l}.$$

The actual post-selection frequency of each group-class profile can then be expressed as

$$u'_{\boldsymbol{g}} = r'_{G,\boldsymbol{g}} \cdot a'_{G,\boldsymbol{g}},$$

where

$$a'_{G,\boldsymbol{g}} = \frac{u'_{\boldsymbol{g}}}{r'_{G,\boldsymbol{g}}}$$

is the post-selection association factor of profile $\boldsymbol{g}$. A value of one indicates that the profile occurs at the frequency expected under independence, whereas values greater or lower than one indicate overrepresentation or underrepresentation, respectively.

The informational change associated with group class $g_l$ at position *l* is the marginal Jeffreys divergence

$$J_{G_l} = \sum_{g_l=1}^{G} \left(p'_{l,g_l} - p_{l,g_l}\right) \log\left(\frac{p'_{l,g_l}}{p_{l,g_l}}\right).$$

The association information generated among the constituent group classes is the Jeffreys divergence between the actual post-selection profile distribution and its fitted independence distribution:

$$J_{G,\text{assoc}} = \sum_{\boldsymbol{g}} \left(u'_{\boldsymbol{g}} - r'_{G,\boldsymbol{g}}\right) \log\left(\frac{u'_{\boldsymbol{g}}}{r'_{G,\boldsymbol{g}}}\right).$$

Consequently, $J_{G,\text{assoc}} \geq 0$ and $J_{G,\text{assoc}} = 0$ if and only if the post-selection distribution of group-class profiles is completely determined by its marginal distributions:

$$u'_{\boldsymbol{g}} = \prod_{l=1}^{L} p'_{l,g_l}.$$

Thus, $J_{G,\text{assoc}}$ quantifies the post-selection association among constituent group classes, measured as the departure of complete group-class profiles from that expected under independence given their post-selection marginal frequencies.

Under the initial-independence assumption, the logarithmic change in the frequency of a group-class profile can be separated as

$$\log\left(\frac{u'_{\boldsymbol{g}}}{u_{\boldsymbol{g}}}\right)=\sum_{l=1}^{L}\log\left(\frac{p'_{l,g_l}}{p_{l,g_l}}\right)+\log\left(a'_{G,\boldsymbol{g}}\right)=\sum_{l=1}^{L}\log\left(\frac{p'_{l,g_l}}{p_{l,g_l}}\right)+\log\left(\frac{u'_{\boldsymbol{g}}}{r'_{G,\boldsymbol{g}}}\right).$$

The first term represents changes in the marginal frequencies of the constituent group classes, whereas the second represents deviation from post-selection independence. Substitution into the set-level Jeffreys information in (3) gives the exact partition

$$J_S=\sum_{l=1}^{L}J_{G_l}+J_{G,\text{assoc}}+E_S,$$

where

$$E_S=\sum_{\boldsymbol{g}}\left(r'_{G,\boldsymbol{g}}-u_{\boldsymbol{g}}\right)\log\left(\frac{u'_{\boldsymbol{g}}}{r'_{G,\boldsymbol{g}}}\right)$$

is the residual term required to complete the partition exactly.

Accordingly, the set-level information $J_S$ is separated into changes in the marginal frequencies of the group classes, information attributable to their post-selection association, and a residual term.

Under the usual local asymptotic approximation, the association divergence can also be used to test independence among the constituent group classes within higher-level sets:

$$n_S\cdot J_{G,\text{assoc}}\sim\chi^2_{\nu_{G,\text{assoc}}},$$

where $n_S$ is the number of independently observed higher-level sets. Each higher-level set contributes one observation of the complete group-class profile $\boldsymbol{g}$; therefore, $n_S$ is not the total number of constituent groups or the number of possible associations. When all $G$ group classes have positive support at each of the $L$ positions and there are no structural constraints on their combinations, the association degrees of freedom are

$$\nu_{G,\text{assoc}}=G^L-1-L\left(G-1\right)$$

More generally, if position $l$ admits $G_l \leq G$ group classes, with no structural constraints on their combinations across positions, the association degrees of freedom become

$$\nu_{G,\mathrm{assoc}} = \prod_{l=1}^{L} G_l - 1 - \sum_{l=1}^{L} \left( G_l - 1 \right)$$

Thus, the hierarchical model distinguishes selection that merely changes the representation of individual group classes from selection that favors or disfavors particular combinations of classes within higher-level sets (Figure 1).

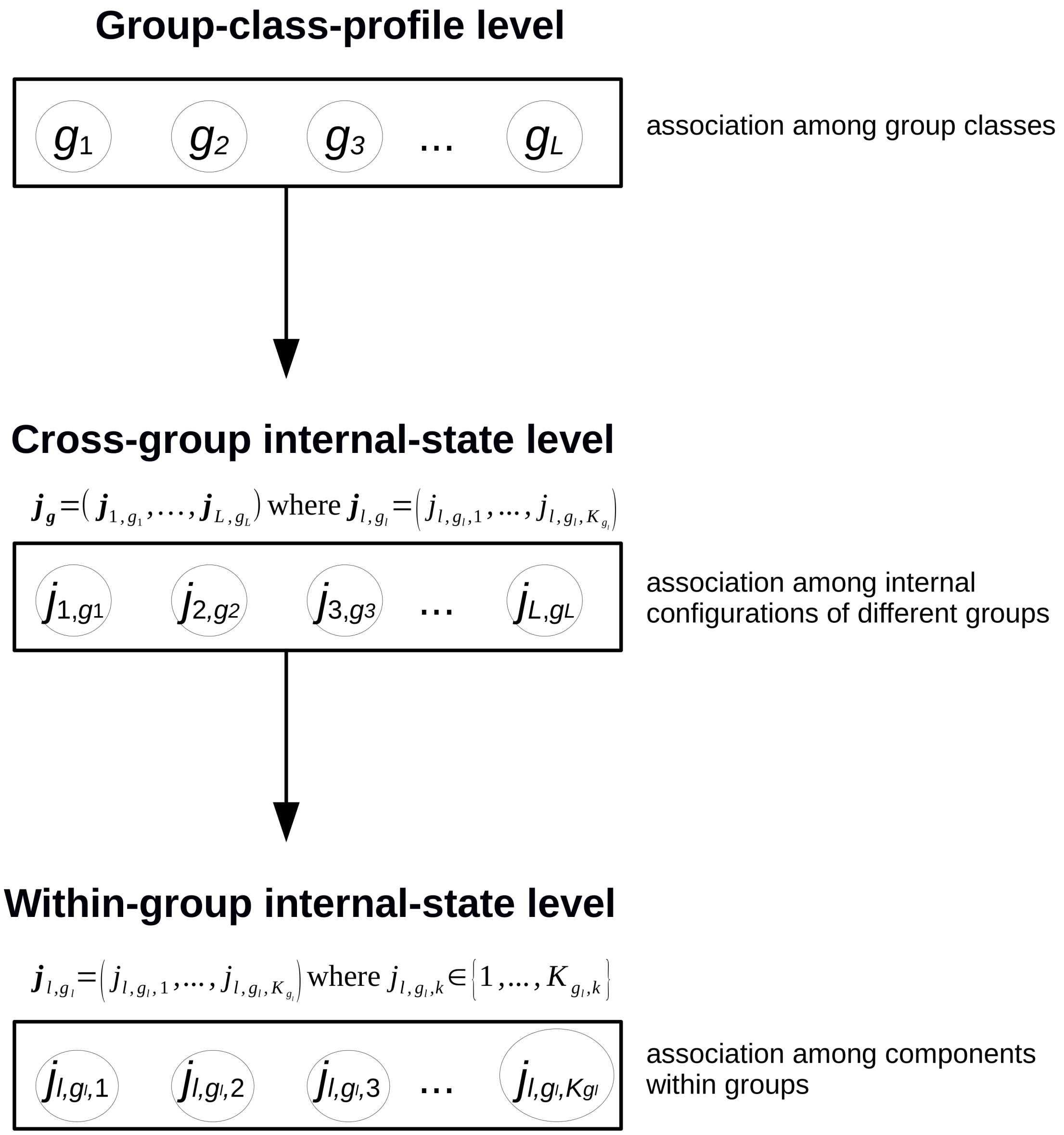

**Figure 1.** Hierarchical multicomponent structure of the Theory of Variable Interactions. Higher-level sets are defined by their group-class profile $\boldsymbol{g}$, each profile contains the internal configurations of its constituent groups $\boldsymbol{j}_{\boldsymbol{g}}$, and each group is further resolved into its component states $j_{l,g_l}$. Informational partitions can therefore be defined at three nested levels: among group classes, among internal configurations of different groups, and among components within groups.

## 4.6 Variable Associations among the Internal Configurations of Different Groups

We next partition the internal information $J_{I,\boldsymbol{g}}$ for each group-class profile into informational changes associated with the internal configurations of its constituent groups at positions $l = 1,\ldots, L$, and an association component among these configurations, together with the residual required for an exact partition. Here, $J_{I,\boldsymbol{g}}$ is the local Jeffreys divergence conditional on a fixed group-class profile $\boldsymbol{g}$; its contribution to the total hierarchical information, including the weighting and the coupling term $E_{SI}$, has already been accounted for in Section 4.4.

For a fixed group-class profile $\boldsymbol{g}$, let

$$h_{\boldsymbol{g},l,\boldsymbol{j}_{l,g_l}}$$

denote the initial frequency of the complete internal configuration $\boldsymbol{j}_{l,g_l}$ of the constituent group occupying position $l$, among higher-level sets sharing profile $\boldsymbol{g}$. For every position $l$,

$$\sum_{\boldsymbol{j}_{l,g_l}} h_{\boldsymbol{g},l,\boldsymbol{j}_{l,g_l}} = 1.$$

Recall that $q_{\boldsymbol{g},\boldsymbol{j}_{\boldsymbol{g}}}$ denotes the frequency of the combined internal configuration $\boldsymbol{j}_{\boldsymbol{g}}$ among higher-level sets sharing group-class profile $\boldsymbol{g}$. We expect that, in the absence of selection favouring particular combinations, the complete internal configurations of the different constituent groups are combined independently within higher-level sets sharing the same group-class profile. Thus,

$$q_{\boldsymbol{g},\boldsymbol{j}_{\boldsymbol{g}}} = \prod_{l=1}^{L} h_{\boldsymbol{g},l,\boldsymbol{j}_{l,g_l}}.$$

Under this expectation, the complete internal configurations observed at different positions may follow distinct marginal distributions across higher-level sets sharing the same group-class profile, while the joint frequency of a particular combination of configurations is determined solely by the product of the corresponding marginal frequencies.

After selection, the joint internal distribution is $q'$. The corresponding marginal frequency of the complete internal configuration of the group occupying position $l$ is obtained by summing over the internal configurations of all other constituent groups:

$$h'_{\boldsymbol{g},l,\boldsymbol{j}_{l,g_l}} = \sum_{\boldsymbol{j}_{\boldsymbol{g},-l}} q'_{\boldsymbol{g},\boldsymbol{j}_{\boldsymbol{g}}},$$

where $\boldsymbol{j}_{\boldsymbol{g},-l}$ denotes the internal configurations of all constituent groups except the group occupying position $l$. The Jeffreys information associated with the complete internal-state distribution of the constituent group occupying position $l$, among higher-level sets sharing group-class profile $\boldsymbol{g}$, is

$$J_{I,\boldsymbol{g},l} = \sum_{\boldsymbol{j}_{l,g_l}} \left(h'_{\boldsymbol{g},l,\boldsymbol{j}_{l,g_l}} - h_{\boldsymbol{g},l,\boldsymbol{j}_{l,g_l}}\right) \log\left(\frac{h'_{\boldsymbol{g},l,\boldsymbol{j}_{l,g_l}}}{h_{\boldsymbol{g},l,\boldsymbol{j}_{l,g_l}}}\right).$$

In the holobiont specialization, each constituent group is itself a holobiont, and $\boldsymbol{j}_{l,g_l}$ specifies its complete host-microbiome configuration. Thus, $h'_{\boldsymbol{g},l,\boldsymbol{j}_{l,g_l}}$ is obtained by summing over the internal configurations of all other holobionts while holding the focal holobiont configuration fixed.

Using the post-selection marginal frequencies, the joint distribution expected under independence is then defined as

$$r'_{B,\boldsymbol{g},\boldsymbol{j}_{\boldsymbol{g}}} = \prod_{l=1}^{L} h'_{\boldsymbol{g},l,\boldsymbol{j}_{l,g_l}}.$$

As above, $r'_B$ preserves the post-selection group-specific internal marginals while removing association among constituent-group configurations. The actual post-selection frequency can therefore be expressed as

$$q'_{\boldsymbol{g},\boldsymbol{j}_{\boldsymbol{g}}} = r'_{B,\boldsymbol{g},\boldsymbol{j}_{\boldsymbol{g}}} \cdot a'_{B,\boldsymbol{g},\boldsymbol{j}_{\boldsymbol{g}}},$$

where

$$a'_{B,\boldsymbol{g},\boldsymbol{j}_{\boldsymbol{g}}} = \frac{q'_{\boldsymbol{g},\boldsymbol{j}_{\boldsymbol{g}}}}{r'_{B,\boldsymbol{g},\boldsymbol{j}_{\boldsymbol{g}}}}$$

is the post-selection association factor among the internal states of the different constituent groups. The subscript $B$ denotes association between groups. Accordingly, $a'_B = 1$ corresponds to independence, whereas deviations from one indicate over- or underrepresentation of particular combined internal configurations.

The corresponding association information is:

$$J_{B,\boldsymbol{g},\text{assoc}} = \sum_{\boldsymbol{j}_g} \left(q'_{\boldsymbol{g},\boldsymbol{j}_g} - r'_{B,\boldsymbol{g},\boldsymbol{j}_g}\right) \log\left(\frac{q'_{\boldsymbol{g},\boldsymbol{j}_g}}{r'_{B,\boldsymbol{g},\boldsymbol{j}_g}}\right).$$

Thus,

$$J_{B,\boldsymbol{g},\text{assoc}} \geq 0,$$

and it is zero if and only if

$$q'_{\boldsymbol{g},\boldsymbol{j}_g} = \prod_{l=1}^{L} h'_{\boldsymbol{g},l,\boldsymbol{j}_{l,g_l}}$$

for every combined internal configuration.

Under the initial-independence assumption,

$$\log\left(\frac{q'_{\boldsymbol{g},\boldsymbol{j}_g}}{q_{\boldsymbol{g},\boldsymbol{j}_g}}\right) = \sum_{l=1}^{L} \log\left(\frac{h'_{\boldsymbol{g},l,\boldsymbol{j}_{l,g_l}}}{h_{\boldsymbol{g},l,\boldsymbol{j}_{l,g_l}}}\right) + \log\left(\frac{q'_{\boldsymbol{g},\boldsymbol{j}_g}}{r'_{B,\boldsymbol{g},\boldsymbol{j}_g}}\right).$$

The exact partition of the internal information among higher-level sets sharing profile $\boldsymbol{g}$ is therefore

$$J_{I,\boldsymbol{g}} = \sum_{l=1}^{L} J_{I,\boldsymbol{g},l} + J_{B,\boldsymbol{g},\text{assoc}} + E_{B,\boldsymbol{g}},$$

where

$$E_{B,\boldsymbol{g}} = \sum_{\boldsymbol{j}_g} \left(r'_{B,\boldsymbol{g},\boldsymbol{j}_g} - q_{\boldsymbol{g},\boldsymbol{j}_g}\right) \log\left(\frac{q'_{\boldsymbol{g},\boldsymbol{j}_g}}{r'_{B,\boldsymbol{g},\boldsymbol{j}_g}}\right)$$

is the residual required to complete the partition exactly.

The association term is a nonnegative Jeffreys divergence, whereas the residual accounts for the finite-change interaction between changes in the group-specific marginals and the association generated among them. This partition therefore detects whether particular internal configurations of one constituent group occur more or less frequently than expected in combination with particular internal configurations of other groups, after the group-class profile has been fixed.

Under the usual local asymptotic approximation, $n_{\boldsymbol{g}}\, J_{B,\boldsymbol{g},assoc}$ follows a chi-squared distribution, where $n_{\mathbf{g}}$ is the number of higher-level sets with profile $\boldsymbol{g}$ (the test is conditional on $n_{\boldsymbol{g}}$, which changes with selection). When all internal configurations have positive support and there are no structural constraints, the degrees of freedom are

$$n_g J_{B,g,\text{assoc}} \approx \chi^2_{\nu_{B,g}}, \quad \nu_{B,g} = \prod_{l=1}^{L} K^c_{g_l} - 1 - \sum_{l=1}^{L} \left( K^c_{g_l} - 1 \right)$$

where $K^c_{g_l} = \prod_k^K g_{l,k}$ is the number of complete internal configurations of a group of class $g_l$.

## 4.7 Variable Associations among Components within Each Group

We finally partition the within-group information $J_{I,\boldsymbol{g},l}$, associated with the constituent group occupying position $l$ within a fixed group-class profile $\boldsymbol{g}$, into informational changes associated with its individual components and an association component among those components, together with the residual required for an exact partition. The complete internal configuration of this constituent group may be resolved into its $K_{g_l}$ internal components. For a fixed group-class profile $\boldsymbol{g}$ and position $l$, let

$$m_{\boldsymbol{g},l,k,j_{l,g_l,k}}$$

denote the initial marginal frequency of state $j_{l,g_l,k}$ of component $k$. For each component

$$\sum_{j_{l,g_l,k}} m_{\boldsymbol{g},l,k,j_{l,g_l,k}} = 1.$$

Under the initial independence assumption, the initial frequency of the complete internal configuration is therefore

$$h_{\boldsymbol{g},l,\boldsymbol{j}_{l,g_l}} = \prod_{k=1}^{K_{g_l}} m_{\boldsymbol{g},l,k,j_{l,g_l,k}}.$$

After selection, the marginal frequency of state $j_{l,g_l,k}$ of component $k$ is obtained by summing over the states of all remaining components:

$$m'_{\boldsymbol{g},l,k,j_{l,g_l,k}} = \sum_{\boldsymbol{j}_{l,g_l,-k}} h'_{\boldsymbol{g},l,\boldsymbol{j}_{l,g_l}},$$

where

$$\boldsymbol{j}_{l,g_l,-k}$$

denotes the internal states of all components of group *l* except component *k*.

The Jeffreys information associated with the marginal distribution of component *k* is

$$J_{\boldsymbol{g},l,k}=\sum_{j_{l,g_l,k}}\left(m'_{\boldsymbol{g},l,k,j_{l,g_l,k}}-m_{\boldsymbol{g},l,k,j_{l,g_l,k}}\right)\log\left(\frac{m'_{\boldsymbol{g},l,k,j_{l,g_l,k}}}{m_{\boldsymbol{g},l,k,j_{l,g_l,k}}}\right).$$

To distinguish marginal changes from changes in association structure, define the post-selection distribution expected under independence as

$$r'_{W,\boldsymbol{g},l,\boldsymbol{j}_{l,g_l}}=\prod_{k=1}^{K_{g_l}}m'_{\boldsymbol{g},l,k,j_{l,g_l,k}}$$

and the corresponding within-group association factor as

$$a'_{W,\boldsymbol{g},l,\boldsymbol{j}_{l,g_l}}=\frac{h'_{\boldsymbol{g},l,\boldsymbol{j}_{l,g_l}}}{r'_{W,\boldsymbol{g},l,\boldsymbol{j}_{l,g_l}}}$$

As at the preceding levels, $r'_W$ preserves the post-selection component marginals while removing their association, and $a'_W = 1$ corresponds to independence.

The association information generated among the components within the group is

$$J_{W,\boldsymbol{g},l,\text{assoc}}=\sum_{\boldsymbol{j}_{l,g_l}}\left(h'_{\boldsymbol{g},l,\boldsymbol{j}_{l,g_l}}-r'_{W,\boldsymbol{g},l,\boldsymbol{j}_{l,g_l}}\right)\log\left(\frac{h'_{\boldsymbol{g},l,\boldsymbol{j}_{l,g_l}}}{r'_{W,\boldsymbol{g},l,\boldsymbol{j}_{l,g_l}}}\right).$$

Thus,

$$J_{W,\boldsymbol{g},l,\text{assoc}}\geq 0,$$

and it is zero if and only if

$$h'_{\boldsymbol{g},l,\boldsymbol{j}_{l,g_l}}=\prod_{k=1}^{K_{g_l}}m'_{\boldsymbol{g},l,k,j_{l,g_l,k}}$$

for every complete internal configuration of group *l*.

Under initial independence,

$$\log\left(\frac{h'_{\boldsymbol{g},l,\boldsymbol{j}_{l,\boldsymbol{g}_l}}}{h_{\boldsymbol{g},l,\boldsymbol{j}_{l,\boldsymbol{g}_l}}}\right)=\sum_{k=1}^{K_{g_l}}\log\left(\frac{m'_{\boldsymbol{g},l,k,j_{l,g_l,k}}}{m_{\boldsymbol{g},l,k,j_{l,g_l,k}}}\right)+\log\left(\frac{h'_{\boldsymbol{g},l,\boldsymbol{j}_{l,\boldsymbol{g}_l}}}{r'_{W,\boldsymbol{g},l,\boldsymbol{j}_{l,\boldsymbol{g}_l}}}\right).$$

The complete internal information of group *l* therefore has the exact partition

$$J_{I,\boldsymbol{g},l}=\sum_{k=1}^{K_{g_l}} J_{\boldsymbol{g},l,k}+J_{W,\boldsymbol{g},l,\text{assoc}}+E_{W,\boldsymbol{g},l},$$

where

$$E_{W,\boldsymbol{g},l}=\sum_{\boldsymbol{j}_{l,\boldsymbol{g}_l}}\left(r'_{W,\boldsymbol{g},l,\boldsymbol{j}_{l,\boldsymbol{g}_l}}-h_{\boldsymbol{g},l,\boldsymbol{j}_{l,\boldsymbol{g}_l}}\right)\log\left(\frac{h'_{\boldsymbol{g},l,\boldsymbol{j}_{l,\boldsymbol{g}_l}}}{r'_{W,\boldsymbol{g},l,\boldsymbol{j}_{l,\boldsymbol{g}_l}}}\right)$$

is the finite-change residual required to complete the partition exactly.

As at the other hierarchical levels, $J_{W,\boldsymbol{g},l,\text{assoc}}$ is a non-negative Jeffreys divergence and can provide the basis for an asymptotic test of independence. In a holobiont specialization, it can test departures from independence among host and microbial components, including host-microbiome associations, as in the holobiont model of Section 2, or associations among microbial components within the holobiont. With $n_{\boldsymbol{g}}$ observations of the group at position *l*, $n_g J_{W,g,l,\text{assoc}}$ follows asymptotically a chi-squared distribution whose degrees of freedom depend only on the local state space,

$$n_g J_{W,g,l,\text{assoc}}\approx\chi^2_{\nu_{W,g,l}},\quad \nu_{W,g,l}=\prod_{k=1}^{K_g} K_{g_l,k}-1-\sum_{k=1}^{K_{g_l}}\left(K_{g_l,k}-1\right).$$

## 4.8 Complete hierarchical informational partition

Substituting the successive marginal-association partitions into the decomposition of the total information gives

$$J_T=\sum_{l=1}^{L} J_{G_l}+J_{G,\text{assoc}}+E_S+\sum_{\boldsymbol{g}}\bar{u}_{\boldsymbol{g}}\left[\sum_{l=1}^{L}\left(\sum_{k=1}^{K_{g_l}} J_{\boldsymbol{g},l,k}+J_{W,\boldsymbol{g},l,\text{assoc}}+E_{W,\boldsymbol{g},l}\right)+J_{B,\boldsymbol{g},\text{assoc}}+E_{B,\boldsymbol{g}}\right]+E_{SI}.$$

The first three terms partition set-level change into marginal group-class information, $J_{G,l}$, association among group classes $J_{G,\text{assoc}}$, and the residual $E_S$. Within each group-class profile $\boldsymbol{g}$, the weighted terms partition internal change into component-level marginal information $J_{\boldsymbol{g},l,k}$, within-group association $J_{W,\boldsymbol{g},l,\text{assoc}}$, and association among the complete internal configurations of different constituent groups $J_{B,\boldsymbol{g},\text{assoc}}$, together with the corresponding residuals $E_{W,\boldsymbol{g},l}$, and $E_{B,\boldsymbol{g}}$. The factors $\bar{u}_{\boldsymbol{g}}$ do not alter the local definition of the lower-level informational terms; they convert information measured conditional on profile $\boldsymbol{g}$ into its contribution to population-wide $J_T$. Accordingly, $E_{SI}$ differs conceptually from $E_S$, $E_{B,\boldsymbol{g}}$, and $E_{W,\boldsymbol{g},l}$: the latter arise from within-level marginal-association partitions, whereas $E_{SI}$ represents coupling between hierarchical levels.

### 4.9 Initial-Independence Convention

The three association levels developed above, also illustrated in Figure 1, rely on the same initial-independence convention, which provides a common reference for distinguishing changes in marginal frequencies from associations generated by selection. Under this convention, the initial joint distribution at each level is determined by the corresponding marginal distributions. Selection may subsequently alter those marginals while preserving independence, or additionally generate non-random associations among the constituent entities. Accordingly, each informational partition separates marginal informational change from a non-negative post-selection association divergence, together with a residual term required to complete the finite-change partition exactly.

This assumption is the same as that adopted in the holobiont models developed above and in the non-random-mating models developed by Carvajal-Rodríguez (2024, 2018). Pre-existing associations and their subsequent strengthening, weakening, or reversal are not considered in the present formulation. The initial-independence convention should be read as the null hypothesis of random assembly from the pool available at each level, with transmission and other frequency-altering processes included in the relative-change function. Departures from that null, at the stage of assembly or at any later stage, are what the association terms quantify.

## 5 Recursive Formulation: A Theory of Variable Interactions

The same reasoning can be applied recursively to an arbitrary number of hierarchical levels. Consider a set at level $n$ whose state is determined by the joint configuration of $K_n$ constituent units:

$$\boldsymbol{x_n} = \left( x_{n,1}, \dots, x_{n,K_n} \right).$$

Let

$$P_{n,\boldsymbol{x_n}}$$

be its joint distribution, and

$$P_{n,k,x_{n,k}}$$

the marginal distribution of constituent $k$. Under the same initial-independence convention, the initial joint distribution at level $n$ factorizes as:

$$P_{n,\boldsymbol{x_n}} = \prod_{k=1}^{K_n} P_{n,k,x_{n,k}}.$$

Using the post-selection marginal distributions, define the fitted independence distribution

$$R'_{n,\boldsymbol{x_n}} = \prod_{k=1}^{K_n} P'_{n,k,x_{n,k}}.$$

The post-selection association factor at level $n$ is defined as

$$a'_{n,\boldsymbol{x_n}} = \frac{P'_{n,\boldsymbol{x_n}}}{R'_{n,\boldsymbol{x_n}}}$$

and hence

$$P'_{n,\boldsymbol{x_n}} = R'_{n,\boldsymbol{x_n}} a'_{n,\boldsymbol{x_n}}.$$

The Jeffreys information associated with the change of the complete set is

$$J_n = \sum_{\boldsymbol{x_n}} \left( P'_{n,\boldsymbol{x_n}} - P_{n,\boldsymbol{x_n}} \right) \log \left( \frac{P'_{n,\boldsymbol{x_n}}}{P_{n,\boldsymbol{x_n}}} \right).$$

For each constituent,

$$J_{n,k} = \sum_{x_{n,k}} \left( P'_{n,k,x_{n,k}} - P_{n,k,x_{n,k}} \right) \log \left( \frac{P'_{n,k,x_{n,k}}}{P_{n,k,x_{n,k}}} \right).$$

The post-selection association information is

$$J_{n,\text{assoc}} = \sum_{x_n} \left(P'_{n,x_n} - R'_{n,x_n}\right) \log\left(\frac{P'_{n,x_n}}{R'_{n,x_n}}\right).$$

The corresponding residual is

$$E_n = \sum_{x_n} \left(R'_{n,x_n} - P_{n,x_n}\right) \log\left(\frac{P'_{n,x_n}}{R'_{n,x_n}}\right).$$

The exact partition is therefore

$$J_n = \sum_{k=1}^{K_n} J_{n,k} + J_{n,\text{assoc}} + E_n .$$

The realized relative fitness at level *n* is correspondingly factorized as

$$\frac{P'_{n,x_n}}{P_{n,x_n}} = a'_{n,x_n} \prod_{k=1}^{K_n} \frac{P'_{n,k,x_{n,k}}}{P_{n,k,x_{n,k}}} .$$

If any constituent *k* is itself a multicomponent set, its informational term $J_{n,k}$ can be partitioned again by the same rule. Repeated application produces a hierarchical informational decomposition extending from elementary components to the highest-level sets. Two related operations should therefore be distinguished in this recursion. Within a fixed hierarchical context, informational change is represented by a local Jeffreys divergence and can be partitioned into marginal, association, and finite-change residual terms. When a lower-level distribution is conditional on a higher-level state whose frequency also changes, however, its contribution to the global Jeffreys information is weighted by the changing frequency of that state, producing an explicit coupling between hierarchical levels. The coupling identity of Section 4.4 holds at every level of the recursion: the local Jeffreys information of each level is weighted by the mean frequency of its higher-level context, and the coupling terms of the successive levels add up (see the Supplementary Mathematical Appendix).

This recursive structure constitutes the Theory of Variable Interactions (TVI): at each local hierarchical level, the Jeffreys information generated by selection, through differential relative fitness among states, is partitioned into changes in the distributions of the constituent entities, association information generated among those entities, and a finite-change residual. More generally, the same replicator-type formulation can accommodate other processes that alter frequencies and associations, as shown by Carvajal-Rodríguez (2026).

In TVI, the degrees of freedom associated with a given local structure are invariant with respect to the hierarchical level at which that structure is embedded; they depend only on its local state-space structure. Hierarchical extension is therefore cumulative in statistical dimensionality: lower-level degrees of freedom are preserved, while each new level introduces additional relational dimensions. Consequently, increasing hierarchical depth enlarges the space of potential informational change, because evolutionary change may be expressed not only through the entities and associations already represented at lower levels, but also through associations emerging at the newly introduced level. These additional contributions may be zero, but they constitute genuinely new dimensions of possible evolutionary change. Because each test is local, its degrees of freedom do not increase with hierarchical depth. They do increase rapidly with the number of constituent units at a given level (for example, $v_{G,\text{assoc}} = G^L - 1 - L(G-1)$), and tests conditional on a context use only the observations that share it, so finer hierarchical partitions reduce the effective sample size per test and increase the number of tests that may require correction for multiplicity.

## 5.1 Recovery of the Preceding Model and Its Specializations

The hierarchical multicomponent-set model developed in Section 4 constitutes the explicit three-level realization of the recursive formulation considered here, once the coupling between levels (Section 4.4 and the Supplementary Mathematical Appendix) is added to the local partition of each level: association may occur among group classes, among the internal configurations of constituent groups, and among components within each group (Figure 1). The preceding model of multicomponent sets under within- and between-group selection (Section 3) is recovered from this hierarchy by setting $L$ = 1, so that each higher-level observational unit contains a single multicomponent group. The population may nevertheless contain many such groups belonging to different group classes, allowing their frequencies to change through between-group selection. Because each observational unit contains only one constituent group, there can be neither association among group classes occupying different positions nor association among the internal states of different constituent groups. Hence,

$$J_{G,\text{assoc}} = E_S = 0$$

and

$$J_{B,\boldsymbol{g},\text{assoc}} = E_{B,\boldsymbol{g}} = 0.$$

The group-class profile reduces to the single index $g$:

$$\boldsymbol{g} = (g),$$

with

$$u_{\boldsymbol{g}} = p_g$$

and the total partition reduces to

$$J_T = J_G + \sum_{g=1}^{G} \bar{p}_g \cdot J_{I,g} + E_{GI},$$

which is the structure derived for multicomponent sets under within- and between-group selection. Within each group class, the initial-independence partition remains

$$J_{I,g} = \sum_{k=1}^{K_g} J_{g,k} + J_{g,assoc} + E_g$$

The tragedy-of-the-commons and holobiont models, being specializations of this recovered model, are consequently also contained within the hierarchical formulation. The hierarchical embedding of the holobiont model, from its internal component structure to holobionts treated as groups and ultimately as constituents of higher-level sets, is summarized in Figure 2.

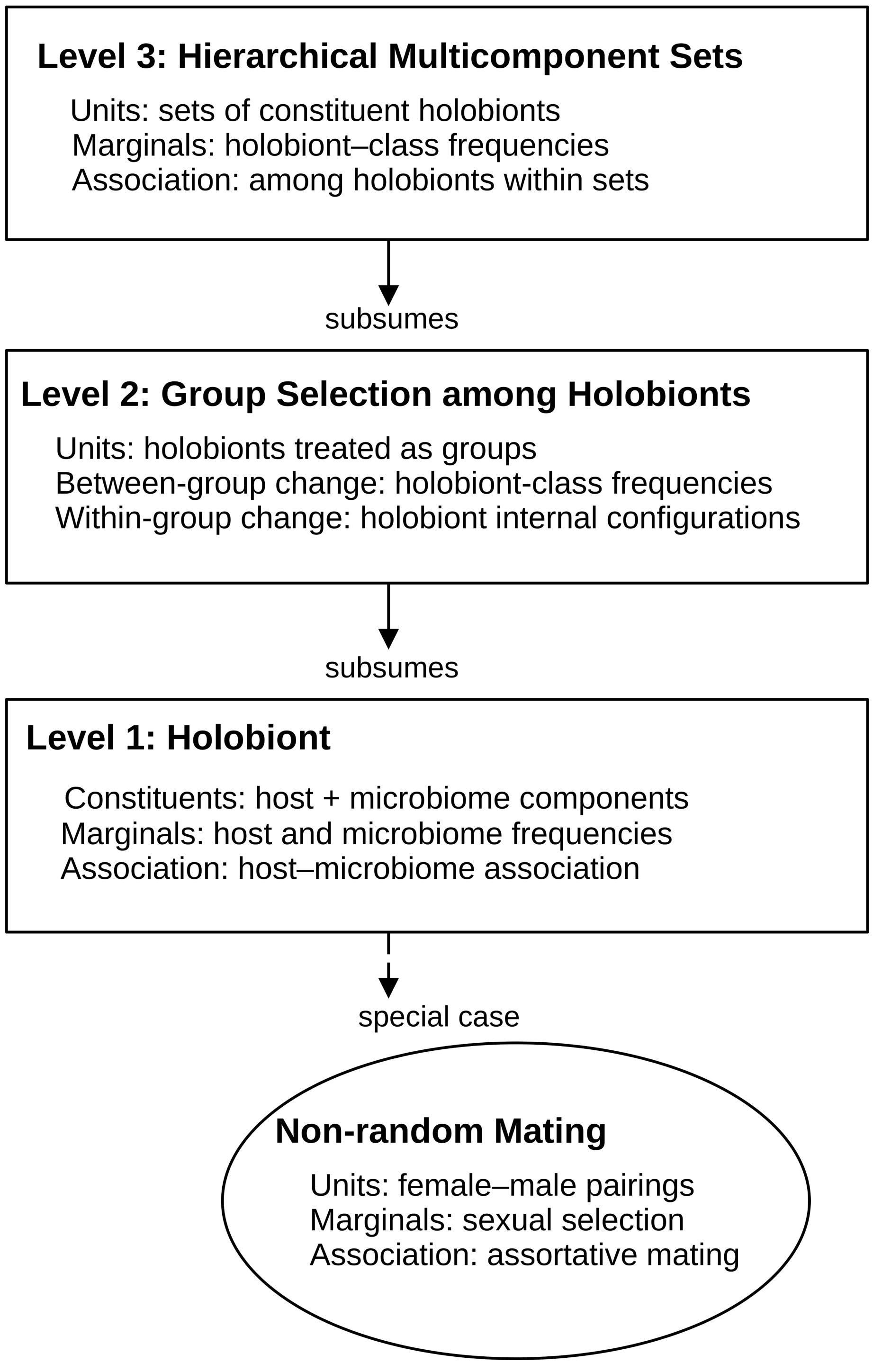

**Figure 2.** Hierarchical relationships among informational models within the Theory of Variable Interactions. At the upper level, hierarchical multicomponent sets contain constituent holobionts and allow associations among them. At the intermediate level, holobionts are treated as groups, distinguishing between-group change in holobiont-class frequencies from within-group change in their internal configurations. At the lowest level, the holobiont model separates changes in host and microbiome marginals from host-microbiome association. Non-random mating is recovered as a special case in which the two components are the two sexes, with marginal and association information corresponding to sexual selection and assortative mating, respectively. Arrows indicate model inclusion. Finite-change residuals are omitted for clarity.

Figure 2 shows that the holobiont admits two partitions of the same total information, depending on how the host is treated. At the lowest level, the host is one of the components, and $J_T = J_H + J_M + J_{\text{assoc}} + E_{\text{holo}}$ (or its multicomponent version, with $E_{\text{multi}}$). At the intermediate level, the holobiont is a group whose class is the host type *i* and whose internal configuration is the microbiome, and $J_T = J_G + \Sigma_i \bar{p}_i J_{I,i} + E_{GI}$, where $J_G$ is the change in host-type frequency (equal to $J_H$) and $J_{I,i}$ measures the change in the microbial distribution conditional on host type *i*, thereby incorporating the effects of both microbial marginal change and host-dependent microbial associations.

$$J_T = J_H + J_M + J_{\text{assoc}} + E_{\text{holo}} = J_G + \sum_i \bar{p}_i \cdot J_{I,i} + E_{GI} .$$

The two expressions are different decompositions of the same quantity: the first is chosen when the question concerns the host-microbiome association, and the second when it concerns the behavior of the microbiome within each host class.

## 6 Discussion

In the present study, we have revised and extended the informational model of the holobiont in both its multicomponent and aggregate formulations. In the multicomponent model, the microbiome may comprise any number of microbial components, each of which can occur in different states, whereas the aggregate model is recovered as the special case in which all microbial components are treated jointly as a single component. For both formulations, we developed statistical tests that distinguish whether observed informational change is attributable to selection acting on the host, on one or more microbial components, or on particular host-microbiome associations. These tests therefore provide a direct means of identifying the hierarchical source of the informational effects generated by selection. In this sense, the present framework provides an information-theoretic contribution to the emerging program of quantitative hologenomics, which has emphasized the need for statistical approaches capable of disentangling host, microbial, and joint effects (Bordenstein and The Holobiont Biology Network, 2024). Our informational partition also provides a complementary perspective to the quantitative-genetic framework recently developed by Week et al. (2025). Whereas this framework partitions additive trait variation into host-genetic, microbial, and gene-microbe covariance components, our approach partitions realized informational change under selection into changes in host and microbial marginal distributions and changes in their association structure. The two approaches therefore address distinct but potentially complementary aspects of host-microbiome evolution: the sources and transmissibility of phenotypic variation on the one hand, and the distributional information generated by evolutionary change on the other. To leading order, $J_T$ equals the variance of relative fitness, that is, the index of the opportunity for selection (Crow, 1958), a relation already established for sexual selection (Carvajal-Rodríguez, 2024; Wade and Arnold, 1980). More generally, $J_T$ is the covariance between relative fitness and its logarithm, the form that Price's (1970) covariance equation takes when the character is log fitness.

We further showed that the informational holobiont model can be embedded within a broader framework, which we call the Theory of Variable Interactions (TVI). By shifting part of the analytical emphasis from biological entities considered in isolation to the associations established among them, TVI belongs to a broader relational tradition in evolutionary thought. It is compatible with the Margulisian emphasis on symbiosis and symbiogenesis as major sources of evolutionary innovation, through which previously autonomous organisms may enter persistent associations and eventually become integrated into new functional wholes (Margulis and Fester, 1991; Sagan, 1967). It also resonates with the relational perspective developed by Lewontin and Levins, in which organisms are simultaneously products and producers of their conditions of existence and evolutionary causation emerges from interactions among genetic, organismal, and environmental factors (Levins and Lewontin, 1985; Lewontin, 2000, 1983, 1974).

Within TVI, this broader relational notion of interaction is represented formally through association structure in joint frequency distributions. From this perspective, an association is not merely a background condition modifying the fitness of otherwise independent entities. Associations may possess structure, specificity, persistence, and evolutionary consequences. They may form, dissolve, become reorganized, or change their functional effects, and in some cases sufficiently stable associations may contribute to the emergence of higher-level biological individuals. TVI does not, however, require every association to constitute an organism, an evolutionary individual, or an independent unit of selection. Its more general claim is that associations can themselves constitute legitimate objects of evolutionary analysis while their constituent entities retain substantial autonomy. In the present formulation, which covers a single selection step with random assembly from the available pool as the reference, this variability is expressed in the post-selection association factor $a'$. Following an association over successive generations would require retaining a non-trivial initial association and partitioning the change in the association factor relative to its initial value, which is left for future work.

This relational emphasis does not diminish the causal importance of genes or the explanatory value of gene-centred evolutionary perspectives. Rather, it places genetic processes within the hierarchical organization of biological systems. Genes, genomes, cells, organisms, symbionts, and groups operate within nested and partially overlapping structures, and their evolutionary effects depend on relationships that may themselves change. In TVI, this distinction is made explicit mathematically: changes in the marginal frequencies of interacting components are separated from changes in their joint distribution. Both contribute to evolutionary information, understood here as structured change in frequency distributions between evolutionary states and quantified using Jeffreys divergence. The resulting partition therefore distinguishes informational change due to changes in the constituent entities from informational change specifically associated with their relational organization.

In this respect, TVI extends the information-handling and holobiont framework developed by Carvajal-Rodríguez (2026) by treating evolutionary change as occurring simultaneously through changes in biological entities and in the structures connecting them. The association parameters of the holobiont model also play a role analogous to the mate-choice parameters estimated in the non-random-mating framework of Carvajal-Rodríguez (2025, 2020). Extending the maximum-likelihood and multimodel estimation procedures developed in that context to host-microbiome configurations represents a promising direction for future work, although it lies beyond the statistical-testing objectives of the present study.

## 6.1 Relation to other interaction-centred approaches

TVI shares with several recent approaches a shift away from treating genes, organisms, populations, or lineages as the only relevant objects of evolutionary explanation. This is evident in the “It’s the Song, Not the Singer” proposal (ITSNTS; Doolittle and Inkpen, 2018), its network-based extension “It Is the Song and the Singers” (ITSATS; Bapteste and Papale, 2021), and the more recent evosystem framework (Papale et al., 2024). These approaches are related to TVI but operate at different explanatory levels.

ITSNTS emphasizes that ecologically relevant processes may persist despite turnover in the biological entities that implement them, whereas ITSATS further reduces the distinction between entities and processes by representing both as patterns of interaction. TVI is compatible with this relational view but addresses a different question. It does not require an interaction or process to persist sufficiently long to constitute a reproduced unit. Transient, newly formed, or dissolving associations may also contribute to evolutionary change. The central question in TVI is instead how much of the informational change observed between evolutionary states is attributable to changes in the interacting components and how much to changes in their joint organization.

The evosystem perspective broadens this relational view to multilevel systems composed of heterogeneous and temporally changing biotic and abiotic interactions (Papale et al., 2024). TVI converges with this perspective in recognizing relational organization as an evolving feature rather than merely an external context. However, their scopes differ. The evosystem framework primarily identifies a broad systemic target of evolutionary explanation, whereas TVI provides a quantitative decomposition of change within a specified system. The categorical Information-Handler framework (Carvajal-Rodríguez, 2026) may provide an additional bridge between these perspectives by representing informationally active entities and their transformations across hierarchical levels, while TVI focuses more specifically on how changes in the relations among such entities contribute to measurable evolutionary information.

These approaches are therefore complementary, although they address different aspects of evolutionary organization. The distinctive contribution of TVI is not simply to recognize the evolutionary importance of interactions, but to provide an explicit, potentially testable informational partition of changes in constituent entities and their associations across hierarchical levels.

A further conceptual connection can be made with the view of the genome as a generative model of the organism (Mitchell and Cheney, 2025) and with related multiscale accounts of genotype-phenotype organization (Hartl and Levin, 2025). These perspectives emphasize that biological outcomes depend not only on constituent elements but also on context-dependent interactions among them, a view compatible with the Information-Handler framework and with the relational emphasis of TVI. Although regulatory and developmental interactions are not formally equivalent to the population-level association distributions considered here, the correspondence may be particularly relevant to holobionts, whose organization can depend jointly on host, microbial, and relational components.

Although the present development focuses primarily on selection, the informational formalism need not be restricted to changes generated by selection. As shown by Carvajal-Rodríguez (2026), evolutionary transitions that can be represented by a normalized replicator-type frequency update can also be expressed in terms of Jeffreys information, with the corresponding relative-change function not necessarily restricted to biological fitness. This opens the possibility of extending TVI to changes in associations generated by processes such as migration, transmission, or other frequency-altering mechanisms. Such extensions would require the relevant process to be explicitly formulated within the same frequency-update framework, but they provide a natural direction for the further development of TVI.

### 6.2 Limitations and future directions

The present formulation has several limits that define the next steps. First, the reference distribution is the independence expected from random assembly from the available pools; associations already present before selection would require retaining an initial association and partitioning its change. Second, the model is deterministic with multinomial sampling; incorporating drift and variance in reproductive success (a stochastic version) is needed to calibrate the tests when drift cannot be neglected. Third, the estimated divergence is biased upwards by approximately $v/n$ under the null hypothesis, where $v$ is the number of degrees of freedom of the term, so it should be bias-corrected when used as a measure of magnitude; this bias does not affect the tests, whose reference distribution already accounts for it. Fourth, the asymptotic chi-squared approximation requires adequate sampling of the configuration space; when many configurations have small expected counts, pooling of classes and Monte Carlo or permutation calibration should be implemented. Fifth, future work could estimate host, microbial, and association parameters by maximum likelihood under appropriately constrained parametric models and compare alternative model structures through multimodel inference. Such models would require identifying constraints to resolve the non-identifiability of the fitness factors $W_H$, $W_M$ and $a_{ij}$ (Section 2.2.1), as in the non-random-mating framework implemented in QInfoMating (Carvajal-Rodríguez, 2025, 2020). Finally, the framework is well suited to designs in which the reference distribution is fixed by design, such as reciprocal transplants or experiments in which hosts are inoculated with defined microbiomes in all host-microbiome combinations, and to systems with well-characterized partners (coral-Symbiodiniaceae, legume-rhizobia, Wolbachia-host).

## CRediT authorship contribution statement

A. Carvajal-Rodríguez: Writing– review & editing, Writing – original draft, Visualization, Validation, Supervision, Methodology, Investigation, Conceptualization.

## Declaration of competing interest

The author declares that they have no known competing financial interests or personal relationships that could have appeared to influence the work reported in this paper.

## Acknowledgements

This work was supported by Xunta de Galicia (Grupo de Referencia Competitiva, ED431C 2024/22), Centro singular de investigación de Galicia accreditation 2024-2027 (ED431G 2023/07) and ERDF A way of making Europe.

## Declaration of generative AI and AI-assisted technologies in the manuscript preparation process.

During the preparation of this work, the author used ChatGPT (-5.6 Sol) and Claude (Sonnet 5.5) to review the style and clarity of the English, as well as to write and format mathematical expressions in LibreOffice Math. All mathematical content and AI-generated text were subsequently reviewed and edited by the author, who assumes full responsibility for the content of the published article.

## References

Bapteste, E., Papale, F., 2021. Modeling the evolution of interconnected processes: It is the song and the singers. BioEssays 43, 2000077. https://doi.org/10.1002/bies.202000077

Bordenstein, S.R., The Holobiont Biology Network, 2024. The disciplinary matrix of holobiont biology. Science 386, 731–732. https://doi.org/10.1126/science.ado2152

Carvajal-Rodríguez, A., 2026. Life as a Categorical Information-Handling System: An Evolutionary Information-Theoretic Model of the Holobiont. Biology 15, 125. https://doi.org/10.3390/biology15020125

Carvajal-Rodríguez, A., 2025. QInfoMating: sexual selection and assortative mating estimation software. BMC Ecology and Evolution 25, 51. https://doi.org/10.1186/s12862-025-02394-8

Carvajal-Rodríguez, A., 2024. Unifying quantification methods for sexual selection and assortative mating using information theory. Theoretical Population Biology 158, 206–215. https://doi.org/10.1016/j.tpb.2024.06.007

Carvajal-Rodríguez, A., 2020. Multi-model inference of non-random mating from an information theoretic approach. Theoretical Population Biology 131, 38–53. https://doi.org/10.1016/j.tpb.2019.11.002

Carvajal-Rodríguez, A., 2018. Non-random mating and information theory. Theoretical Population Biology 120, 103–113. https://doi.org/10.1016/j.tpb.2018.01.003

Crow, J.F., 1958. Some possibilities for measuring selection intensities in man. Hum Biol 30, 1–13.

Doolittle, W.F., Inkpen, S.A., 2018. Processes and patterns of interaction as units of selection: An introduction to ITSNTS thinking. Proceedings of the National Academy of Sciences 115, 4006–4014. https://doi.org/10.1073/pnas.1722232115

Frank, S.A., 2025. Natural selection at multiple scales. Evolution 79, 1166–1184. https://doi.org/10.1093/evolut/qpaf037

Frank, S.A., 2012. Natural selection. V. How to read the fundamental equations of evolutionary change in terms of information theory. J Evol Biol 25, 2377–96. https://doi.org/10.1111/jeb.12010

Hartl, B., Levin, M., 2025. What does evolution make? Learning in living lineages and machines. Trends in Genetics 41, 480–496. https://doi.org/10.1016/j.tig.2025.04.002

Kullback, S., 1997. Information Theory and Statistics, New edition. ed. Dover Publications, Mineola, N.Y.

Lancaster, H.O., 1951. Complex Contingency Tables Treated by the Partition of χ2. Royal Statistical Society. Journal. Series B: Methodological 13, 242–249. https://doi.org/10.1111/j.2517-6161.1951.tb00089.x

Levins, R., Lewontin, R., 1985. The Dialectical Biologist. Harvard University Press.

Lewontin, R.C., 2000. The Triple Helix: Gene, Organism, and Environment. Harvard University Press.

Lewontin, R.C., 1983. The Organism as the Subject and Object of Evolution. Scientia 77, 65.

Lewontin, R.C., 1974. The genetic basis of evolutionary change. Columbia University Press, New York.

Margulis, L., Fester, R. (Eds.), 1991. Symbiogenesis and symbionticism, in: Symbiosis as a Source of Evolutionary Innovation: Speciation and Morphogenesis. MIT Press, Cambridge, MA, USA, pp. 1–14.

McGill, W.J., 1954. Multivariate information transmission. Psychometrika 19, 97–116. https://doi.org/10.1007/BF02289159

Mitchell, K.J., Cheney, N., 2025. The Genomic Code: the genome instantiates a generative model of the organism. Trends in Genetics 41, 462–479. https://doi.org/10.1016/j.tig.2025.01.008

Papale, F., Not, F., Bapteste, É., Haraoui, L.-P., 2024. The evosystem: A centerpiece for evolutionary studies. BioEssays 46, 2300169. https://doi.org/10.1002/bies.202300169

Price, G.R., 1970. Selection and covariance. Nature 227, 520–521.

Roughgarden, J., 2020. Holobiont Evolution: Mathematical Model with Vertical vs. Horizontal Microbiome Transmission. Philosophy, Theory, and Practice in Biology 12. https://doi.org/10.3998/ptpbio.16039257.0012.002

Sagan, L., 1967. On the origin of mitosing cells. Journal of Theoretical Biology 14, 225-IN6. https://doi.org/10.1016/0022-5193(67)90079-3

Theis, K.R., Dheilly, N.M., Klassen, J.L., Brucker, R.M., Baines, J.F., Bosch, T.C.G., Cryan, J.F., Gilbert, S.F., Goodnight, C.J., Lloyd, E.A., Sapp, J., Vandenkoornhuyse, P., Zilber-Rosenberg, I., Rosenberg, E., Bordenstein, S.R., 2016. Getting the Hologenome Concept Right: an Eco-Evolutionary Framework for Hosts and Their Microbiomes. mSystems 1, e00028-16. https://doi.org/10.1128/mSystems.00028-16

Wade, M.J., Arnold, S.J., 1980. The intensity of sexual selection in relation to male sexual behaviour, female choice, and sperm precedence. Animal Behaviour 28, 446–461. https://doi.org/10.1016/S0003-3472(80)80052-2

Watanabe, S., 1960. Information Theoretical Analysis of Multivariate Correlation. IBM Journal of Research and Development 4, 66–82. https://doi.org/10.1147/rd.41.0066

Week, B., Ralph, P.L., Tavalire, H.F., Cresko, W.A., Bohannan, B.J.M., 2025. Quantitative genetics of microbiome-mediated traits. Evol 79, 2487–2502. https://doi.org/10.1093/evolut/qpaf171

Zilber-Rosenberg, I., Rosenberg, E., 2008. Role of microorganisms in the evolution of animals and plants: the hologenome theory of evolution. FEMS Microbiol Rev 32, 723–735. https://doi.org/10.1111/j.1574-6976.2008.00123.x

## Supplementary Mathematical Appendix

## Multicomponent Sets Under Within- and Between-Group Selection

Consider a population structured into $G$ group classes, indexed by $g$. The frequency of group class $g$ is denoted by $p_g$. The simplest nontrivial case consists of two group classes, although the formulation below applies to any G ≥ 1.

A group belonging to group class $g$ contains $K_g \geq 1$ components, indexed by $k$. Component $k$ can occur in $K_{g,k} \geq 1$ possible states, indexed by $j_{g,k}$. Consequently, the number of possible internal configurations associated with group class $g$ is

$$C_g = \prod_{k=1}^{K_g} K_{g,k}$$

A complete internal configuration is represented by the vector

$$\boldsymbol{j_g} = \left( j_{g,1}, \dots, j_{g,K_g} \right)$$

and $q_{g,\boldsymbol{j}_g}$ denotes the frequency of configuration $j_g$ among groups belonging to group class $g$. Thus,

$$\sum_{\boldsymbol{j}_g} q_{g,\boldsymbol{j}_g} = 1$$

for every $g$.

Selection can act simultaneously at two levels. First, internal configurations may differ in their relative success within their respective group class. Second, group classes may differ in their collective success relative to other group classes. Accordingly, the relative fitness of a complete state can be factorized as

$$\omega_{g,\boldsymbol{j}_g} = \omega_{I,g,\boldsymbol{j}_g} \cdot \omega_{G,g}$$

where $\omega_{I,g,j_g}$ denotes the relative fitness of internal configuration $\boldsymbol{j}_g$ within group class $g$, and $\omega_{G,g}$ denotes the relative collective fitness of group class $g$.

The two relative fitnesses are normalized at their respective levels:

$$\sum_{\boldsymbol{j}_g} q_{g,\boldsymbol{j}_g} \omega_{I,g,\boldsymbol{j}_g} = 1$$

for every $g$, and

$$\sum_{g=1}^{G} p_g \omega_{G,g} = 1$$

The frequencies of the group classes and of their internal configurations therefore change according to

$$p'_g = p_g \omega_{G,g}$$

and

$$q'_{g,\boldsymbol{j}_g} = q_{g,\boldsymbol{j}_g} \omega_{I,g,\boldsymbol{j}_g}$$

respectively. Between-group selection changes the representation of the different group classes, whereas within-group selection changes the distribution of internal configurations among groups belonging to each group class.

Because

$$\omega_{G,g} = \frac{p'_g}{p_g}$$

the mean change in the group-level character $\log(\omega_{G,g})$, caused by differential collective fitness among group classes, is

$$J_G = \sum_{g=1}^{G} (p'_g - p_g) \log\left(\frac{p'_g}{p_g}\right)$$

Thus, $J_G$ is the informational change generated by between-group selection.

Similarly, for each group class $g$, the informational change generated at the internal, or within-group-class, level is

$$J_{I,g} = \sum_{\boldsymbol{j}_g} (q'_{g,\boldsymbol{j}_g} - q_{g,\boldsymbol{j}_g}) \log\left(\frac{q'_{g,\boldsymbol{j}_g}}{q_{g,\boldsymbol{j}_g}}\right)$$

## Partition of the within-group-class information

Because the internal state of group class $g$ is multicomponent, $J_{I,g}$ can itself be decomposed into changes in the marginal distributions of the individual components, changes in their joint organization, and a residual term.

For component $k$, its initial marginal frequency is obtained by holding $j_{g,k}$ fixed and summing over all possible states of the remaining components:

$$m_{g,k,j_{g,k}} = \sum_{j_{g,1}=1}^{K_{g,1}} \cdots \sum_{j_{g,k-1}=1}^{K_{g,k-1}} \sum_{j_{g,k+1}=1}^{K_{g,k+1}} \cdots \sum_{j_{g,K_g}=1}^{K_{g,K_g}} q_{g,\boldsymbol{j_g}}$$

Similarly, after within-group selection,

$$m'_{g,k,j_{g,k}} = \sum_{j_{g,1}=1}^{K_{g,1}} \cdots \sum_{j_{g,k-1}=1}^{K_{g,k-1}} \sum_{j_{g,k+1}=1}^{K_{g,k+1}} \cdots \sum_{j_{g,K_g}=1}^{K_{g,K_g}} q'_{g,\boldsymbol{j_g}}$$

Both marginal distributions are normalized:

$$\sum_{j_{g,k}=1}^{K_{g,k}} m_{g,k,j_{g,k}} = 1$$

and

$$\sum_{j_{g,k}=1}^{K_{g,k}} m'_{g,k,j_{g,k}} = 1$$

The informational change in the marginal distribution of component $k$ is therefore

$$J_{g,k} = \sum_{j_{g,k}=1}^{K_{g,k}} \left( m'_{g,k,j_{g,k}} - m_{g,k,j_{g,k}} \right) \log \left( \frac{m'_{g,k,j_{g,k}}}{m_{g,k,j_{g,k}}} \right)$$

To identify informational change attributable to associations among components, define the distribution expected under independence from the post-selection marginal frequencies as

$$r_{g,\boldsymbol{j_g}} = \prod_{k=1}^{K_g} m'_{g,k,j_{g,k}}$$

The distribution $r_{g,\boldsymbol{j}_g}$ has exactly the same component marginal frequencies as $q'_{g,\boldsymbol{j}_g}$, but contains no statistical associations among components.

The association term is the Jeffreys divergence between these two distributions:

$$J_{g,\text{assoc}} = \sum_{\boldsymbol{j}_g} \left( q'_{g,\boldsymbol{j}_g} - r_{g,\boldsymbol{j}_g} \right) \log \left( \frac{q'_{g,\boldsymbol{j}_g}}{r_{g,\boldsymbol{j}_g}} \right)$$

It follows that $J_{g,\text{assoc}}$ is zero when the post-selection joint distribution is completely determined by the product of its component marginals, and positive when particular combinations of component states occur more or less frequently than expected under independence.

For the partition below, assume that the initial internal configuration distribution is itself independent across components:

$$q_{g,\boldsymbol{j}_g} = \prod_{k=1}^{K_g} m_{g,k,j_{g,k}}$$

Under this assumption, the Jeffreys divergence between the two product distributions $r_{g,\boldsymbol{j}g}$ *and* $q_{g,\boldsymbol{j}_g}$ decomposes into the sum of the component divergences:

$$\sum_{k=1}^{K_g} J_{g,k} = \sum_{\boldsymbol{j}_g} \left( r_{g,\boldsymbol{j}_g} - q_{g,\boldsymbol{j}_g} \right) \log \left( \frac{r_{g,\boldsymbol{j}_g}}{q_{g,\boldsymbol{j}_g}} \right)$$

Moreover, because $q'_{g,\boldsymbol{j}_g}$ *and* $r_{g,\boldsymbol{j}_g}$ have the same post-selection marginal distributions,

$$\sum_{\boldsymbol{j}_g} \left( q'_{g,\boldsymbol{j}_g} - r_{g,\boldsymbol{j}_g} \right) \log \left( \frac{r_{g,\boldsymbol{j}_g}}{q_{g,\boldsymbol{j}_g}} \right) = 0$$

The reason is that, under initial independence, the logarithmic ratio in the preceding expression separates into a sum of functions depending on one component at a time, and the expectation of each such function is identical under *q*' and *r* because both distributions possess the same marginals.

Now write

$$\log \left( \frac{q'_{g,\boldsymbol{j}_g}}{q_{g,\boldsymbol{j}_g}} \right) = \log \left( \frac{q'_{g,\boldsymbol{j}_g}}{r_{g,\boldsymbol{j}_g}} \right) + \log \left( \frac{r_{g,\boldsymbol{j}_g}}{q_{g,\boldsymbol{j}_g}} \right)$$

and define

$$E_g = \sum_{\boldsymbol{j}_g} \left( r_{g,\boldsymbol{j}_g} - q_{g,\boldsymbol{j}_g} \right) \log \left( \frac{q'_{g,\boldsymbol{j}_g}}{r_{g,\boldsymbol{j}_g}} \right)$$

Then the exact partition is

$$J_{I,g}=\left(\sum_{k=1}^{K_g} J_{g,k}\right)+J_{g,\mathrm{assoc}}+E_g$$

Unlike the Jeffreys terms, $E_g$ is not itself a divergence and may therefore be positive, negative, or zero. It is the residual required for exact additivity of the component-marginal and association contributions.

Thus, within-group selection may change the frequencies of the states belonging to individual components, the statistical associations among those components, or both.

## Total Informational Change Across Group Classes and Internal Configurations

The preceding quantities distinguish informational change occurring between group classes from that occurring within each group class. The total informational change can alternatively be evaluated directly from the joint distribution of group-class identity and internal configuration.

Define the global frequency of the state consisting of group class g and internal configuration $\boldsymbol{j}_g$ as

$$F_{g,\boldsymbol{j}_g}=p_g \cdot q_{g,\boldsymbol{j}_g}$$

After selection,

$$F'_{g,\boldsymbol{j}_g}=p'_g \cdot q'_{g,\boldsymbol{j}_g}$$

Because both internal distributions are normalized,

$$\sum_{\boldsymbol{j}_g} q_{g,\boldsymbol{j}_g}=\sum_{\boldsymbol{j}_g} q'_{g,\boldsymbol{j}_g}=1$$

and because the relative fitness factors are normalized at their respective levels, their product is a globally normalized relative fitness:

$$\omega_{g,\boldsymbol{j}_g}=\omega_{G,g} \cdot \omega_{I,g,\boldsymbol{j}_g}$$

Consequently,

$$F'_{g,\boldsymbol{j}_g}=F_{g,\boldsymbol{j}_g} \cdot \omega_{g,\boldsymbol{j}_g}$$

The total informational change generated by the complete transformation is therefore

$$J_T = \sum_{g=1}^{G} \sum_{\mathbf{j}_g} \left( F'_{g,\mathbf{j}_g} - F_{g,\mathbf{j}_g} \right) \log \left( \frac{F'_{g,\mathbf{j}_g}}{F_{g,\mathbf{j}_g}} \right)$$

Substituting the product representation of the global frequencies gives

$$J_T = \sum_{g=1}^{G} \sum_{\mathbf{j}_g} \left( p'_g q'_{g,\mathbf{j}_g} - p_g q_{g,\mathbf{j}_g} \right) \log \left( \frac{p'_g q'_{g,\mathbf{j}_g}}{p_g q_{g,\mathbf{j}_g}} \right)$$

The logarithm separates exactly into its between-group and internal components:

$$\log \left( \frac{p'_g q'_{g,\mathbf{j}_g}}{p_g q_{g,\mathbf{j}_g}} \right) = \log \left( \frac{p'_g}{p_g} \right) + \log \left( \frac{q'_{g,\mathbf{j}_g}}{q_{g,\mathbf{j}_g}} \right)$$

Therefore,

$$J_T = T_G + T_I$$

where

$$T_G = \sum_{g=1}^{G} \sum_{\mathbf{j}_g} \left( p'_g q'_{g,\mathbf{j}_g} - p_g q_{g,\mathbf{j}_g} \right) \log \left( \frac{p'_g}{p_g} \right)$$

and

$$T_I = \sum_{g=1}^{G} \sum_{\mathbf{j}_g} \left( p'_g q'_{g,\mathbf{j}_g} - p_g q_{g,\mathbf{j}_g} \right) \log \left( \frac{q'_{g,\mathbf{j}_g}}{q_{g,\mathbf{j}_g}} \right)$$

Because the group-class frequency ratio is independent of internal configuration,

$$T_G = \sum_{g=1}^{G} \log \left( \frac{p'_g}{p_g} \right) \left[ p'_g \sum_{\mathbf{j}_g} q'_{g,\mathbf{j}_g} - p_g \sum_{\mathbf{j}_g} q_{g,\mathbf{j}_g} \right]$$

Normalization of the internal distributions then gives

$$T_G = \sum_{g=1}^{G} \left( p'_g - p_g \right) \log \left( \frac{p'_g}{p_g} \right)$$

and hence

$$T_G = J_G$$

For the internal part,

$$T_I=\sum_{g=1}^{G}\left[p'_g\sum_{j_g} q'_{g,j_g}\log\left(\frac{q'_{g,j_g}}{q_{g,j_g}}\right)-p_g\sum_{j_g} q_{g,j_g}\log\left(\frac{q'_{g,j_g}}{q_{g,j_g}}\right)\right]$$

Define the forward internal Kullback–Leibler (KL) divergence as

$$D^{+}_{I,g}=\sum_{j_g} q'_{g,j_g}\log\left(\frac{q'_{g,j_g}}{q_{g,j_g}}\right)$$

and the reverse divergence as

$$D^{-}_{I,g}=\sum_{j_g} q_{g,j_g}\log\left(\frac{q_{g,j_g}}{q'_{g,j_g}}\right)$$

It follows immediately that

$$T_I=\sum_{g=1}^{G}\left(p'_g\cdot D^{+}_{I,g}+p_g\cdot D^{-}_{I,g}\right)$$

The total information therefore has the exact partition

$$J_T=J_G+\sum_{g=1}^{G}\left(p'_g\cdot D^{+}_{I,g}+p_g\cdot D^{-}_{I,g}\right)$$

For each group class,

$$J_{I,g}=D^{+}_{I,g}+D^{-}_{I,g}$$

However, the internal contribution to $J_T$ cannot in general be written simply as a group-frequency-weighted sum of the $J_{I,g}$, because its forward and reverse components receive different weights. The forward divergence is weighted by the post-selection frequency $p'_g$, whereas the reverse divergence is weighted by the initial frequency $p_g$.

Define the mean group-class frequency

$$\bar{p}_g=\frac{p_g+p'_g}{2}$$

Then

$$p'_g\cdot D^{+}_{I,g}+p_g\cdot D^{-}_{I,g}=\bar{p}_g\cdot J_{I,g}+\frac{p'_g-p_g}{2}\cdot\left(D^{+}_{I,g}-D^{-}_{I,g}\right)$$

and consequently

$$J_T = J_G + \sum_{g=1}^{G} \bar{p}_g \cdot J_{I,g} + E_{GI}$$

where the between-within coupling term is

$$E_{GI} = \frac{1}{2} \sum_{g=1}^{G} \left(p'_g - p_g\right) \cdot \left(D^{+}_{I,g} - D^{-}_{I,g}\right)$$

This term arises because changes in the representation of group classes alter the weights assigned to the two directional components of their internal informational change.

### *Properties of the Between-Within Coupling Term*

For convenience, define the directional asymmetry of the internal informational change in group class *g* as

$$A_g = D^{+}_{I,g} - D^{-}_{I,g}$$

The coupling term then becomes

$$E_{GI} = \frac{1}{2} \sum_{g=1}^{G} \left(p'_g - p_g\right) \cdot A_g$$

Therefore, the necessary and sufficient condition for its disappearance is

$$\sum_{g=1}^{G} \left(p'_g - p_g\right) \cdot A_g = 0$$

The coupling term is not itself a divergence and may be positive, negative, or zero.

### *Covariance representation*

Because between-group selection satisfies

$$p'_g = p_g \, \omega_{G,g}$$

the coupling term can also be written as

$$E_{GI} = \frac{1}{2} \sum_{g=1}^{G} p_g \left(\omega_{G,g} - 1\right) \cdot A_g$$

Since collective relative fitness is normalized,

$$\sum_{g=1}^{G} p_g \omega_{G,g} = 1$$

this expression is exactly one half of the covariance, taken with respect to the initial group-class distribution, between collective relative fitness and internal directional asymmetry:

$$E_{GI} = \frac{1}{2} \operatorname{Cov}_p \left( \omega_{G,g}, A_g \right)$$

Thus, $E_{GI}$ measures the association between the collective relative fitness of a group class and the directional asymmetry of the informational transformation occurring internally within that class.

### *No change in group-class frequencies*

If

$$p'_g = p_g$$

for every $g$, then

$$E_{GI} = 0$$

and

$$J_G = 0$$

Consequently,

$$J_T = \sum_{g=1}^{G} p_g \cdot J_{I,g}$$

### *No internal change*

If the internal distribution of a particular group class remains unchanged,

$$q'_{g,\boldsymbol{j}_g} = q_{g,\boldsymbol{j}_g}$$

for every configuration $\boldsymbol{j}_g$, then

$$D^{+}_{I,g} = D^{-}_{I,g} = 0$$

and therefore

$$A_g = 0$$

Such a group class may change in frequency through between-group selection without contributing to $E_{GI}$.

## Equality of the directional internal divergences

More generally, a group class makes no contribution to $E_{GI}$ whenever

$$D^{+}_{I,g} = D^{-}_{I,g}$$

even when its internal distribution changes.

The exact condition for this equality is

$$\sum_{\boldsymbol{j}_g} \left( q'_{g,\boldsymbol{j}_g} + q_{g,\boldsymbol{j}_g} \right) \log \left( \frac{q'_{g,\boldsymbol{j}_g}}{q_{g,\boldsymbol{j}_g}} \right) = 0$$

Using the internal replicator equation,

$$q'_{g,\boldsymbol{j}_g} = q_{g,\boldsymbol{j}_g} \cdot \omega_{I,g,\boldsymbol{j}_g}$$

the same condition becomes

$$\sum_{\boldsymbol{j}_g} q_{g,\boldsymbol{j}_g} \cdot \left( 1 + \omega_{I,g,\boldsymbol{j}_g} \right) \cdot \log \left( \omega_{I,g,\boldsymbol{j}_g} \right) = 0$$

This condition does not imply absence of within-group selection. Rather, it means that the directional asymmetry of the transformation vanishes, so that its forward and reverse KL divergences have equal magnitude.

A simple sufficient example is a symmetric exchange of two internal frequencies:

$$q_g = (a, b)$$

and

$$q'_g = (b, a)$$

with a ≠ b. More generally, equality also holds when the post-selection distribution is obtained by exchanging frequencies in disjoint pairs, possibly leaving other configurations unchanged:

$$q'_{g,\boldsymbol{j_g}} = q_{g,\pi(\boldsymbol{j_g})}$$

with

$$\pi(\pi(\boldsymbol{j_g})) = \boldsymbol{j_g}.$$

*Homogeneous directional asymmetry among group classes*

Suppose that

$$A_g = A$$

for every g. Since

$$\sum_{g=1}^{G} (p'_g - p_g) = 0$$

it follows that

$$E_{GI} = 0$$

This case may involve both between-group and within-group selection. The coupling disappears because the same internal directional asymmetry is shared by all group classes and is therefore uncorrelated with their changes in frequency.

*Compensation among group classes*

The contribution of every group class need not vanish separately. The coupling term also disappears whenever positive and negative contributions compensate:

$$\sum_{g=1}^{G} (p'_g - p_g) \cdot A_g = 0$$

Thus, $E_{GI} = 0$ can occur even when group-class frequencies and internal configuration distributions both change.

For two group classes,

$$p'_2 - p_2 = -(p'_1 - p_1)$$

and therefore

$$E_{GI} = \frac{1}{2}\left(p'_1 - p_1\right)\cdot\left(A_1 - A_2\right)$$

Therefore, $E_{GI} = 0$ if and only if either the group-class frequencies remain unchanged or the two group classes have the same directional asymmetry. In the latter case, $E_{GI}$ may vanish even though both between-group and within-group changes occur.

### *Weak within-group selection*

Under weak internal change, write

$$q'_{g,j_g} = q_{g,j_g} + \delta_{g,j_g}$$

with

$$\sum_{j_g} \delta_{g,j_g} = 0$$

and with the absolute changes small relative to the corresponding initial frequencies.

Taylor expansion of the directional KL divergences gives

$$D^{+}_{I,g} - D^{-}_{I,g} = \frac{1}{6}\sum_{j_g} \frac{\delta^3_{g,j_g}}{q^2_{g,j_q}} + O\left(\delta^4\right)$$

Thus, the forward and reverse divergences agree through second order:

$$D^{+}_{I,g} \approx D^{-}_{I,g}$$

and consequently

$$E_{GI} \approx 0$$

under sufficiently weak internal change.

The total informational change is then approximately

$$J_T \approx J_G + \sum_{g=1}^{G} \bar{p}_g \cdot J_{I,g}$$

If changes in group-class frequencies and internal configuration frequencies are both small and of comparable magnitude, $E_{GI}$ is of higher order than the principal between-group and within-group Jeffreys terms.

# Special Cases and Recovery of Previous Models

## The Tragedy-of-the-Commons Model

The general model reduces to a single-component internal system when

$$K_g = 1$$

In that case no association among internal components can exist, and

$$J_{I,g} = J_{g,1}$$

To connect this formulation with the tragedy-of-the-commons model presented by Frank (2025) and subsequently expressed in informational terms by Carvajal-Rodríguez (2026), let $q$, in this subsection, denote the initial frequency of a focal competitive type within a group. Thus, $q$ acts as the label of a group composition and should not be confused with the general internal-frequency notation $q_{g,j_g}$.

Let $p_q$ denote the frequency of groups in which the focal type has frequency $q$ and the alternative type has frequency 1 - $q$, that is, groups with internal composition ($q$, 1 - $q$), hereafter referred to simply as composition $q$.

If $y_q$ is the mean competitive tendency of groups with composition $q$, their normalized collective fitness is

$$\omega_{G,q} = \frac{\left(\kappa - y_q\right)^s}{\bar{w}_G}$$

where $s$ measures the relative intensity of between-group selection and

$$\bar{w}_G = \sum_q p_q \left(\kappa - y_q\right)^s$$

The frequency of groups with composition $q$ then changes according to

$$p'_q = p_q \, \omega_{G,q}$$

and the corresponding between-group information is

$$J_G=\sum_q (p'_q - p_q)\log\left(\frac{p'_q}{p_q}\right)$$

Within groups of composition $q$, let $f_{x,q}$ denote the frequency of individuals or entities with competitive tendency $x$. The mean competitive tendency is

$$y_q=\sum_x f_{x,q}\,x$$

After intragroup competition,

$$f'_{x,q}=f_{x,q}\frac{x}{y_q}$$

and the informational change generated by within-group selection is

$$J_{I,q}=\sum_x (f'_{x,q}-f_{x,q})\log\left(\frac{f'_{x,q}}{f_{x,q}}\right)$$

The global state is specified by group composition $q$ and competitive tendency $x$. Over a single selection step, $q$ labels the group class defined by its initial composition. The corresponding initial and post-selection global frequencies are

$$F_{q,x}=p_q\cdot f_{x,q}$$

and

$$F'_{q,x}=p'_q\cdot f'_{x,q}$$

Therefore,

$$J_T=\sum_q\sum_x (F'_{q,x}-F_{q,x})\log\left(\frac{F'_{q,x}}{F_{q,x}}\right)$$

Using the general result derived above,

$$J_T=J_G+\sum_q \left(p'_q\cdot D^{+}_{I,q}+p_q\cdot D^{-}_{I,q}\right)$$

where

$$D^{+}_{I,q}=\sum_x f'_{x,q}\log\left(\frac{f'_{x,q}}{f_{x,q}}\right)$$

and

$$D^{-}_{I,q}=\sum_x f_{x,q}\log\left(\frac{f_{x,q}}{f'_{x,q}}\right)$$

with

$$J_{I,q}=D^{+}_{I,q}+D^{-}_{I,q}$$

Equivalently,

$$J_T=J_G+\sum_q \bar{p}_q\cdot J_{I,q}+E_{GI}$$

where

$$\bar{p}_q=\frac{p_q+p'_q}{2}$$

and

$$E_{GI}=\frac{1}{2}\sum_q\left(p'_q-p_q\right)\cdot\left(D^{+}_{I,q}-D^{-}_{I,q}\right)$$

The covariance identity derived for the general model becomes

$$E_{GI}=\frac{1}{2}\,\mathrm{Cov}_p\left(\omega_{G,q},D^{+}_{I,q}-D^{-}_{I,q}\right)$$

Thus, Frank's within-group and group-level fitness components correspond to the informational quantities $J_{I,q}$ and $J_G$, respectively, whereas their combined effect across the complete structured population is represented by $J_T$. The additional term $E_{GI}$ quantifies the coupling between collective relative fitness and the directional asymmetry of the within-group informational transformation.

## The Multicomponent Holobiont Model

The multicomponent holobiont model is recovered by considering a single group class, $G = 1$, with

$$p_1=p'_1=1$$

and

$$\omega_{G,1}=1$$

Therefore,

$$J_G = 0$$

and all informational change occurs at the internal level:

$$J_T = J_{I,1}$$

The internal state of this single group class represents the complete holobiont and comprises one host component and $K$ microbial components. Hence,

$$K_1 = K + 1$$

Component 1 represents the host and has $K_H$ possible types:

$$K_{1,1} = K_H$$

Components 2,...,$K$ + 1 represent the microbial components. Microbial component $M_k$, has $K_{M_k}$ possible states

$$K_{1,k+1} = K_{M_k}$$

for $k$ = 1,...,$K$.

The complete internal configuration can therefore be relabelled as

$$\boldsymbol{j_1} = (i, j_1, \ldots, j_K)$$

where $i$ denotes the host type and $j_k$ the state of microbial component $M_k$.

The joint frequencies are identified as

$$q_{1,\boldsymbol{j_1}} = q_{i, j_1, \ldots, j_K}$$

and

$$q'_{1,\boldsymbol{j_1}} = q'_{i, j_1, \ldots, j_K}$$

The marginal frequencies of component 1 correspond to the host-type frequencies:

$$m_{1,1,i} = p_i$$

and

$$m'_{1,1,i} = p'_i$$

For microbial component $M_k$,

$$m_{1,k+1,j_k} = m_{k,j_k}$$

and

$$m'_{1,k+1,j_k} = m'_{k,j_k}$$

Under the initial-independence assumption,

$$q_{i,j_1,\dots,j_K} = p_i \prod_{k=1}^{K} m_{k,j_k}$$

The corresponding post-selection independence reference is

$$r_{i,j_1,\dots,j_K} = p'_i \prod_{k=1}^{K} m'_{k,j_k}$$

The association term becomes

$$J_{assoc} = \sum_{i,j_1,\dots,j_K} \left(q'_{i,j_1,\dots,j_K} - r_{i,j_1,\dots,j_K}\right) \log\left(\frac{q'_{i,j_1,\dots,j_K}}{r_{i,j_1,\dots,j_K}}\right)$$

and the residual term is

$$E_{multi} = \sum_{i,j_1,\dots,j_K} \left(r_{i,j_1,\dots,j_K} - q_{i,j_1,\dots,j_K}\right) \log\left(\frac{q'_{i,j_1,\dots,j_K}}{r_{i,j_1,\dots,j_K}}\right)$$

Using the identifications

$$J_{1,1} = J_H, J_{1,k+1} = J_{M_k}, J_{1,\mathrm{assoc}} = J_{assoc}$$

and

$$E_1 = E_{multi}$$

the general internal partition becomes

$$J_T = \left(J_H + \sum_{k=1}^{K} J_{M_k}\right) + J_{assoc} + E_{multi}$$

Thus, the multicomponent holobiont is an exact special case of the general multicomponent-set model,

obtained by treating the complete holobiont as the internal state of a single group class.

## Decomposition of the Association Term into Host-Component Contributions

In this subsection, $J_{\text{assoc}}$ denotes the association term of the multicomponent holobiont model, which measures the departure of $q'$ from the independence reference $r_{i,j} = p'_i \prod_k m'_{k,j_k}$. To identify which microbial components are associated with the host, let $q'_{i,j_k}$ be the post-selection frequency of host type $i$ and state $j_k$ of component $M_k$, obtained by summing $q'_{i,j}$ over the states of all the other components. Because $r$ is a product, its corresponding marginal is $p'_i\, m'_{k,j_k}$, and the host-component association term is

$$q'_{i,j_k} = \sum_{\mathbf{j}_{-k}} q'_{i,\mathbf{j}}$$

$$J_{\text{assoc},k} = \sum_{i,j_k} \left(q'_{i,j_k} - p'_i m'_{k,j_k}\right) \log \frac{q'_{i,j_k}}{p'_i m'_{k,j_k}}$$

### *Pairwise-separable fitness model*

Suppose that the relative fitness factorizes into a host factor and one factor per microbial component, each of which may depend on the host type but on no other component:

$$\omega_{i,\mathbf{j}} = \omega_i \prod_{k=1}^{K} \omega^{(k)}_{i,j_k}$$

Under initial independence, $q'_{i,j} = q_{i,j}\,\omega_{i,j} = p_i \omega_i \prod_k m_{k,j_k} \omega^{(k)}_{i,j_k}$. For fixed $i$, this expression factorizes into terms that each depend on a single $j_k$, so the microbial components are conditionally independent given the host type after selection. In this case the global association term decomposes exactly into the host-component terms,

$$J_{\text{assoc}} = \sum_{k=1}^{K} J_{\text{assoc},k}$$

*General case and conditional remainder*

Without separability, the sum of the host-component terms need not exhaust $J_{\text{assoc}}$. Let $q'(j \mid i) = q'_{i,j} / p'_{\text{i}}$ and $q'(j_k \mid i) = q'_{i,jk} / p'_i$ be the conditional distributions given host type $i$, and let $C$ be the ratio between the joint conditional distribution of the microbial states and the product of its conditional marginals,

$$C_{i,\boldsymbol{j}} = \frac{q'(\boldsymbol{j}|i)}{\prod_{k=1}^{K} q'(j_k|i)}$$

Then $q' / r$ equals the product of the host-component ratios times $C$,

$$\frac{q'_{i,\boldsymbol{j}}}{r_{i,\boldsymbol{j}}} = \left[\frac{\prod_{k=1}^{K} q'_{i,j_k}}{p'_i m'_{k,j_k}}\right] C_{i,\boldsymbol{j}}$$

Taking logarithms, multiplying by $(q' - r)$ and summing, the $k$-th logarithmic term depends only on $(i, j_k)$, so its sum equals $J_{\text{assoc},k}$. Therefore

$$J_{\text{assoc}} = \sum_{k=1}^{K} J_{\text{assoc},k} + R_{\text{cond}}, \qquad R_{\text{cond}} = \sum_{i,\boldsymbol{j}} (q'_{i,\boldsymbol{j}} - r_{i,\boldsymbol{j}}) \log C_{i,\boldsymbol{j}}$$

If the microbial components are independent among themselves within each host type, then $C = 1$ and $R_{\text{cond}} = 0$, which recovers the separable case. $R_{\text{cond}}$ is a signed residual arising from conditional dependence among microbial components given the host type. A non-zero $R_{\text{cond}}$ necessarily implies conditional dependence, whereas $R_{\text{cond}} = 0$ does not guarantee conditional independence, since contributions may cancel. Thus, $R_{\text{cond}}$ provides a sufficient indication of conditional dependence when non-zero, but is not a complete measure of it. It does not measure marginal dependence among components, nor the host-component associations, which are already included in the $J_{\text{assoc},k}$ terms. Writing $R_{\text{cond}}$ as the difference between two sums,

$$R_{\text{cond}} = \sum_{i} p'_i \text{TC}_i - \sum_{i,\boldsymbol{j}} r_{i,\boldsymbol{j}} \log C_{i,\boldsymbol{j}}, \qquad \text{TC}_i = \sum_{\boldsymbol{j}} q'(\boldsymbol{j}|i) \log C_{i,\boldsymbol{j}} \geq 0$$

where $\text{TC}_{\text{i}}$ is the total correlation of the microbial components within host type $i$. The first sum is non-negative, but the second is a cross term with no definite sign. Hence $R_{\text{cond}}$ is not a divergence and may be positive, negative, or zero, as is the case for $E_{\text{multi}}$. The two remainders are, however, different quantities: $E_{\text{multi}}$ is the residual of the partition of $J_T$, whereas $R_{\text{cond}}$ is internal to $J_{\text{assoc}}$. Together,

$$J_T = J_H + \sum_{k=1}^{K} J_{Mk} + \sum_{k=1}^{K} J_{\text{assoc},k} + R_{\text{cond}} + E_{\text{multi}}$$

*Testing conditional dependence among microbial components*

The pairwise terms are computed on tables collapsed over the other components: $J_{\text{assoc},k}$ uses only $q'_{i,j_k}$, that is, the distribution of $M_k$ within each host type, and discards how the states of different components combine within a host type. Consequently, in the non-separable case a component $M_k$ can appear associated with the host merely because it depends on another component that is associated with the host, and the dependence among components within host types is not captured by the pairwise terms. Because $R_{\text{cond}}$ is not a divergence, no test is associated with it. The conditional dependence is instead tested through the likelihood-ratio statistic of conditional independence of the microbial components given the host type,

$$G_{\text{cond}} = 2 \sum_{i=1}^{K_H} n_i \sum_{\boldsymbol{j}} q'(\boldsymbol{j}|i) \log C_{i,\boldsymbol{j}} \approx \chi^2_{K_H \nu_{M,\text{assoc}}}, \quad \nu_{M,\text{assoc}} = K_M - 1 - \sum_{k=1}^{K} (K_{Mk} - 1), \quad K_M = \prod_{k=1}^{K} K_{Mk}$$

where $n_i$ is the number of holobionts of host type $i$ in the sample and the frequencies are the sample estimates. The association degrees of freedom of the multicomponent model,

$$\nu_{assoc} = K_H (K_M - 1) - \Sigma_k (K_{Mk} - 1),$$

split into two parts. The first is the sum of the degrees of freedom of the $K$ host-component tests, $\Sigma_k \nu_{assoc,k}$, where $\nu_{assoc,k} = (K_H - 1)(K_{M_k} - 1)$ is that of the host × $M_k$ table. The second, $K_H \nu_{M,\text{assoc}}$, is what remains, so that

$$\nu_{\text{assoc}} = \sum_{k=1}^{K} \nu_{\text{assoc},k} + K_H \nu_{M,\text{assoc}}, \quad \nu_{\text{assoc},k} = (K_H - 1)(K_{M_k} - 1)$$

Hence, $G_{\text{cond}}$ tests conditional independence among microbial components given the host type, corresponding to the $K_H \nu_{M,\text{assoc}}$ degrees of freedom not represented by the host-component association tests. However, $G_{\text{cond}}$ is not, in general, equal to $nR_{\text{cond}}$, because $R_{\text{cond}}$ is a signed residual of the Jeffreys partition, whereas $G_{\text{cond}}$ is a likelihood-ratio statistic.

### *Illustration*

In the toy example of the main text, the fitness table is a product over components,

$$\omega_{i,j_1,j_2} = f_i(j_1) f_i(j_2), \quad f_S(A) = \frac{4}{3}, \quad f_S(B) = \frac{2}{3}, \quad f_L(A) = \frac{2}{3}, \quad f_L(B) = \frac{4}{3}, \quad f_M(A) = f_M(B) = 1$$

where S, M and L are the host types and A and B the states of components $M_1$ and $M_2$. The example therefore satisfies the pairwise-separable model, and the host–component terms add up to the global association term (Table S1): $J_{\text{assoc},1} + J_{\text{assoc},2} = 0.154033 = J_{\text{assoc}}$, and $R_{\text{cond}} = 0$. The degrees of freedom are consistent: the global association term has $K_H(K_M - 1) - \Sigma_k(K_{M_k} - 1) = 3 \times 3 - 2 = 7$ degrees of freedom, which equals 2 + 2 for the two host-component terms plus $K_H\, \nu_{M,\text{assoc}} = 3$ for the remainder. The marginal association between $M_1$ and $M_2$ is also significant ($nJ = 5.94$, 1 degree of freedom, $P = 0.015$; odds ratio 1.35, obtained by collapsing the post-selection table over host type), but it is induced by the host: within each host type the odds ratio is exactly 1, so $R_{cond} = 0$ and $G_{\text{cond}} = 0$.

**Table S1. Decomposition of the association divergence $J_{\text{assoc}}$ into host-component terms in the octopus holobiont example; the pairwise-separable model holds, so the conditional remainder $R_{\text{cond}}$ is zero.**

| Informational term | Interpretation | Value | Test statistic | df | *p*-value |
|---|---|---|---|---|---|
| $J_{\text{assoc},1}$ | Host × $M_1$ association | 0.077016 | $nJ_{\text{assoc},1} = 83.18$ | (3−1)(2−1) = 2 | $8.7 \times 10^{-19}$ |
| $J_{\text{assoc},2}$ | Host × $M_2$ association | 0.077016 | $\text{nJ}_{\text{assoc},2} = 83.18$ | 2 | $8.7 \times 10^{-19}$ |
| $R_{\text{cond}}$ | Dependence between $M_1$ and $M_2$ within host types | 0 | $G_{\text{cond}} = 0$ | $K_H \nu_{M,\text{assoc}} = 3 \times 1 = 3$ | 1 |

### *A non-separable example*

Consider $n = 600$ holobionts with two host types and the same two microbial components $M_1$ and $M_2$ (states A and B), with reference frequencies 1/2 for each host type and for each microbial state, so that the expected count of each configuration under random association is 75. The observed counts are given in Table S2.

**Table S2. Observed counts.**

| Host type | $M_{1,A}$ $M_{2,A}$ | $M_{1,A}$ $M_{2,B}$ | $M_{1,B}$ $M_{2,A}$ | $M_{1,B}$ $M_{2,B}$ | Total |
|---|---|---|---|---|---|
| Host 1 | 150 | 50 | 50 | 50 | 300 |
| Host 2 | 40 | 90 | 110 | 60 | 300 |
| Total | 190 | 140 | 160 | 110 | 600 |

In host 1 the association between $M_1$ and $M_2$ is positive (odds ratio 3.0, log odds ratio 1.10), whereas in host 2 it is negative (odds ratio 0.24, log odds ratio −1.42), so the fitness table cannot be written as a product of component effects and the pairwise-separable model does not hold. The marginal frequencies are 300/600 for the hosts, 330/270 for $M_1$ (A/B) and 350/250 for $M_2$ (A/B). The informational partition is

$$J_T = J_H + J_{M_1} + J_{M_2} + J_{\text{assoc}} + E_{\text{multi}} = 0 + 0.010034 + 0.028039 + 0.182360 - 0.013959 = 0.206473$$

and, within the association term,

$$J_{\text{assoc}} = J_{\text{assoc},1} + J_{\text{assoc},2} + R_{\text{cond}} = 0.056082 + 0.028881 + 0.097396 = 0.182360 .$$

Table S3 gives the tests. Because both host types have the same frequency, $J_H = 0$. Neither remainder is zero: $E_{\text{multi}} = -0.013959$ and $R_{\text{cond}} = 0.097396$. In this example the pairwise terms capture less than half (47 %) of the association information, 0.084963 out of 0.182360. The remainder $R_{\text{cond}}$ is the difference between a non-negative part, the total correlation within hosts weighted by $p'_i$, equal to 0.044172 (0.030575 in host 1 and 0.057768 in host 2), and a cross term equal to −0.053224:

$$R_{\text{cond}} = \sum_i p'_i \, \mathrm{TC}_i - \sum_{i,j} r_{i,j} \log C_{i,j} = 0.044172 - (-0.053224) = 0.097396 .$$

The direct test of conditional independence gives $G_{\text{cond}} = 53.0$ (2 degrees of freedom, $P = 3.1 \times 10^{-12}$), in agreement with the conclusion that the dependence between the two components is conditional on the host and opposite in sign between hosts.

An additional analysis can be performed by collapsing Table S2 over host types to examine the marginal association between $M_1$ and $M_2$, irrespective of the host. The resulting 2 × 2 table has observed counts (AA, AB, BA, BB) = (190, 140, 160, 110), with expected counts under independence of (192.5, 137.5, 157.5, 112.5), obtained from the observed microbial marginal frequencies. The Jeffreys divergence between these observed and expected distributions is 0.000289 ($nJ$ = 0.173, df = 1, $P$ = 0.677), indicating no significant marginal association between $M_1$ and $M_2$. This contrasts with the strong conditional dependence detected within host types ($G_{\text{cond}}$ = 53.0, df = 2, $P = 3.1 \times 10^{-12}$), where the odds ratios are 3.0 and 0.24 for hosts 1 and 2, respectively. Thus, aggregating over host types largely masks the opposing microbial associations present within each host type. This marginal divergence is a separate analysis and does not enter the informational partition of $J_T$ reported in Table S3.

**Table S3. Total and partitioned informational change in the non-separable example.**

| Informational term | Value | Test statistic | df | P-value |
|---|---|---|---|---|
| $J_T$ | 0.206473 | $nJ_T = 123.88$ | 7 | $1.2 \times 10^{-23}$ |
| $J_H$ | 0 | $nJ_H = 0$ | 1 | 1 |
| $J_{M1}$ | 0.010034 | $nJ_{M1} = 6.02$ | 1 | 0.014 |
| $J_{M2}$ | 0.028039 | $nJ_{M2} = 16.82$ | 1 | $4.1 \times 10^{-5}$ |
| $J_{\text{assoc}}$ | 0.182360 | $nJ_{\text{assoc}} = 109.42$ | 4 | $9.7 \times 10^{-23}$ |
| $J_{\text{assoc},1}$ | 0.056082 | $nJ_{\text{assoc},1} = 33.65$ | 1 | $6.6 \times 10^{-9}$ |
| $J_{\text{assoc},2}$ | 0.028881 | $nJ_{\text{assoc},2} = 17.33$ | 1 | $3.1 \times 10^{-5}$ |
| $R_{\text{cond}}$ | 0.097396 | $G_{\text{cond}} = 53.0$ | 2 | $3.1 \times 10^{-12}$ |
| $E_{\text{multi}}$ | −0.013959 | — | — | — |

## The Aggregate Holobiont Model

The aggregate holobiont model is obtained by replacing the $K$ microbial components by a single component representing the microbiome as a whole. The single group class then contains two components:

$$K_1 = 2$$

Component 1 represents the host and component 2 the aggregate microbiome:

$$K_{1,1} = K_H$$

and

$$K_{1,2}=K_M$$

Let $q_{i,j}$ denote the joint frequency of host type $i$ and aggregate microbiome type $j$. The host and microbiome marginals are

$$p_i=\sum_{j=1}^{K_M} q_{i,j}$$

and

$$m_j=\sum_{i=1}^{K_H} q_{i,j}$$

with corresponding post-selection marginals $p'_i$ and $m'_j$.

Under initial independence,

$$q_{i,j}=p_i m_j$$

and the post-selection independence reference is

$$r_{i,j}=p'_i m'_j$$

The aggregate association term is therefore

$$J_{assoc}=\sum_{i=1}^{K_H}\sum_{j=1}^{K_M}\left(q'_{i,j}-r_{i,j}\right)\log\left(\frac{q'_{i,j}}{r_{i,j}}\right)$$

whereas the corresponding residual is

$$E_{holo}=\sum_{i=1}^{K_H}\sum_{j=1}^{K_M}\left(r_{i,j}-q_{i,j}\right)\log\left(\frac{q'_{i,j}}{r_{i,j}}\right)$$

The aggregate partition is consequently

$$J_T=J_H+J_M+J_{\mathrm{assoc}}+E_{\mathrm{holo}}$$

Here $J_M$ measures informational change in the marginal distribution of aggregate microbiome states, whereas the association term measures departure from independence between host type and aggregate microbiome state.

The same symbol $J_{\mathrm{assoc}}$ is used in the aggregate and multicomponent formulations because it plays the same formal role, although the underlying independence reference differs: in the aggregate model it concerns association between host and whole-microbiome classes, whereas in the multicomponent model it concerns the complete joint organization of host and individual microbial components.

### *Exact relabelling of the multicomponent microbiome*

An exact aggregation is obtained if every complete microbial configuration is treated as one distinct state of the aggregate microbiome. The number of aggregate states is then

$$K_M = \prod_{k=1}^{K} K_{M_k}$$

and there is a one-to-one correspondence

$$j \Leftrightarrow (j_1, \dots, j_K)$$

The joint frequencies are therefore merely relabelled:

$$q_{i,j} = q_{i,j_1,\dots,j_K}$$

and

$$q'_{i,j} = q'_{i,j_1,\dots,j_K}$$

Consequently, the total Jeffreys information $J_T$ is exactly preserved.

This does not imply term-by-term equivalence of the aggregate and multicomponent partitions. In the aggregate representation,

$$J_M = \sum_{j=1}^{K_M} (m'_j - m_j) \log\left(\frac{m'_j}{m_j}\right)$$

measures change in the distribution of complete microbial configurations. By contrast, each $J_{M_k}$ measures only change in the marginal distribution of microbial component $k$. Therefore, in general,

$$J_M \neq \sum_{k=1}^{K} J_{M_k}$$

because $J_M$ may contain informational change associated with changes in the joint organization of the microbial components.

Equality holds when the marginal joint distribution of the microbial components, after marginalizing over host types, factorizes into the product of the microbial-component marginals both before and after selection. Independence of microbial components within each host type is not by itself sufficient, because associations between host types and microbial states may induce statistical associations among microbial components after marginalization over the host.

$$J_M = \sum_{k=1}^{K} J_{M_k}$$

By contrast, if several distinct microbial configurations are combined into a single aggregate microbiome state, the mapping is many-to-one and constitutes a genuine coarse-graining. Information distinguishing the collapsed configurations is then discarded, and the resulting total informational change cannot exceed that obtained from the fully resolved multicomponent representation.

Although exact relabelling preserves the complete joint distribution and therefore $J_T$, it removes the component-wise resolution of the informational partition. A genuinely coarser aggregate representation may be useful when reducing dimensionality or sparse sampling is desirable, but this comes at the cost of losing information about differential changes in individual microbial components and their specific contribution to joint holobiont structure.

## Hierarchical Multicomponent Sets: A Theory of Variable Interactions

The preceding formulation separates selection between group classes from selection acting on multicomponent configurations within those classes. The same construction can be iterated when the entities forming a higher-level set are themselves structured groups. This produces a nested hierarchy in which changes in marginal frequencies and changes in association can occur at several organizational levels.

### Hierarchical notation

Consider a higher-level set containing $L$ constituent groups, indexed by $l$. Constituent group $l$ belongs to group class $g_l$. The complete profile of group classes defining the higher-level set is

$$\boldsymbol{g}=(g_1,\dots,g_L)$$

Let $u_{\boldsymbol{g}}$ denote the frequency of higher-level sets having group-class profile $\boldsymbol{g}$, and let $p_{l,g_l}$ denote the corresponding marginal frequency of class $g_l$ at position or constituent-group category $l$.

Under independence among the constituent group classes before selection,

$$u_{\boldsymbol{g}}=\prod_{l=1}^{L} p_{l,g_l}$$

After selection, define the association factor at the group-class-profile level as

$$a'_{G,\boldsymbol{g}}=\frac{u'_{\boldsymbol{g}}}{\prod_{l=1}^{L} p'_{l,g_l}}$$

Thus, $a'_{G,\boldsymbol{g}}$ measures the post-selection departure of the observed group-class profile from that expected from the product of its post-selection marginal frequencies.

Because the initial association factor is one under the assumed initial independence,

$$\frac{u'_{\boldsymbol{g}}}{u_{\boldsymbol{g}}}=a'_{G,\boldsymbol{g}}\prod_{l=1}^{L}\frac{p'_{l,g_l}}{p_{l,g_l}}$$

This gives the first level of the hierarchical factorization.

## Associations among the internal states of constituent groups

Each constituent group $l$ of class $g_l$ may itself occur in different internal configurations, represented by the vector

$$\boldsymbol{j}_{\boldsymbol{l},\boldsymbol{g}_l}=(j_{l,g_l,1},\dots,j_{l,g_l,K_{g_l}})$$

The combined internal configuration of all constituent groups within a higher-level set of profile $\boldsymbol{g}$ is

$$\boldsymbol{j}_{\boldsymbol{g}}=(\boldsymbol{j}_{\boldsymbol{1},\boldsymbol{g}_1},\dots,\boldsymbol{j}_{\boldsymbol{L},\boldsymbol{g}_L})$$

Let $q_{\boldsymbol{g},\boldsymbol{j}\boldsymbol{g}}$ denote the frequency of complete internal profile $\boldsymbol{j}_{\boldsymbol{g}}$ among higher-level sets with group-class profile $\boldsymbol{g}$.

Let

$$h_{\boldsymbol{g},l,\boldsymbol{j}_{l,g_l}}$$

denote the marginal frequency of internal configuration $\boldsymbol{j}_{l,g_l}$ for constituent group *l* within higher-level sets of profile ***g***.

Under initial independence among the internal states of the constituent groups,

$$q_{\boldsymbol{g},\boldsymbol{j}_g}=\prod_{l=1}^{L} h_{\boldsymbol{g},l,\boldsymbol{j}_{l,g_l}}$$

Define the post-selection association factor among constituent-group internal states as

$$a'_{B,\boldsymbol{g},\boldsymbol{j}_g}=\frac{q'_{\boldsymbol{g},\boldsymbol{j}_g}}{\prod_{l=1}^{L} h'_{\boldsymbol{g},l,\boldsymbol{j}_{l,g_l}}}$$

It follows that

$$\frac{q'_{\boldsymbol{g},\boldsymbol{j}_g}}{q_{\boldsymbol{g},\boldsymbol{j}_g}}=a'_{B,\boldsymbol{g},\boldsymbol{j}_g}\prod_{l=1}^{L}\frac{h'_{\boldsymbol{g},l,\boldsymbol{j}_{l,g_l}}}{h_{\boldsymbol{g},l,\boldsymbol{j}_{l,g_l}}}$$

The factor $a'_{B,\boldsymbol{g},\boldsymbol{j}_g}$ therefore represents association generated among the internal states of the different constituent groups within the higher-level set.

## Associations among components within each constituent group

A constituent group *l* belonging to class $g_l$ may itself contain $K_{g_l}$ components. Its internal configuration is

$$\boldsymbol{j}_{l,g_l}=\left(j_{l,g_l,1},\dots,j_{l,g_l,K_{g_l}}\right)$$

Let $m_{\boldsymbol{g},l,k,j_{l,g_l,k}}$ denote the marginal frequency of state $j_{l,g_l,k}$ of component *k* within constituent group *l*.

Under initial independence among the components of that constituent group,

$$h_{\boldsymbol{g},l,\boldsymbol{j}_{l,g_l}}=\prod_{k=1}^{K_{g_l}} m_{\boldsymbol{g},l,k,j_{l,g_l,k}}$$

Define the corresponding post-selection within-group association factor as

$$a'_{W,\boldsymbol{g},l,\boldsymbol{j}_{l,g_l}} = \frac{h'_{\boldsymbol{g},l,\boldsymbol{j}_{l,g_l}}}{\prod_{k=1}^{K_{g_l}} m'_{\boldsymbol{g},l,k,j_{l,g_l,k}}}$$

Therefore,

$$\frac{h'_{\boldsymbol{g},l,\boldsymbol{j}_{l,g_l}}}{h_{\boldsymbol{g},l,\boldsymbol{j}_{l,g_l}}} = a'_{W,\boldsymbol{g},l,\boldsymbol{j}_{l,g_l}} \prod_{k=1}^{K_{g_l}} \frac{m'_{\boldsymbol{g},l,k,j_{l,g_l,k}}}{m_{\boldsymbol{g},l,k,j_{l,g_l,k}}}$$

## Factorization of realized relative fitness across the hierarchy

The global frequency of a complete hierarchical state is

$$F_{\boldsymbol{g},\boldsymbol{j}_g} = u_{\boldsymbol{g}} \cdot q_{\boldsymbol{g},\boldsymbol{j}_g}$$

and after selection,

$$F'_{\boldsymbol{g},\boldsymbol{j}_g} = u'_{\boldsymbol{g}} \cdot q'_{\boldsymbol{g},\boldsymbol{j}_g}$$

Hence, its realized relative fitness is

$$\omega_{\boldsymbol{g},\boldsymbol{j}_g} = \frac{F'_{\boldsymbol{g},\boldsymbol{j}_g}}{F_{\boldsymbol{g},\boldsymbol{j}_g}}$$

Substituting the successive hierarchical factorizations gives

$$\frac{F'_{\boldsymbol{g},\boldsymbol{j}_g}}{F_{\boldsymbol{g},\boldsymbol{j}_g}} = a'_{G,\boldsymbol{g}} a'_{B,\boldsymbol{g},\boldsymbol{j}_g} \prod_{l=1}^{L} \left[ \frac{p'_{l,g_l}}{p_{l,g_l}} a'_{W,\boldsymbol{g},l,\boldsymbol{j}_{l,g_l}} \prod_{k=1}^{K_{g_l}} \frac{m'_{\boldsymbol{g},l,k,j_{l,g_l,k}}}{m_{\boldsymbol{g},l,k,j_{l,g_l,k}}} \right]$$

Thus, realized relative fitness across the complete hierarchy is decomposed into changes in marginal frequencies together with association factors operating at distinct organizational levels.

The factor $a'_{G,\boldsymbol{g}}$ represents association among the group classes composing the higher-level set; $a'_{B,\boldsymbol{g},\boldsymbol{j}_g}$ represents association among the internal states of its constituent groups; and $a'_{W,\boldsymbol{g},l,\boldsymbol{j}_{l,g_l}}$ represents association among components within constituent group $l$.

These association factors quantify the part of realized relative fitness that cannot be represented by changes in the corresponding marginal frequencies alone. Because the association structure itself may change under selection, they provide a formal representation of variable interactions at successive hierarchical levels.

The hierarchical formulation therefore extends the same principle used throughout the preceding models. Evolutionary change can alter both the representation of the constituent entities and the way in which those entities are associated. Iterating this decomposition across organizational levels provides the mathematical basis for a Theory of Variable Interactions.

## Hierarchical Coupling Identity at Any Depth

Let a hierarchical state be $x = (x_1, \ldots, x_n)$, with $x_1$ the highest level, and let $\boldsymbol{c_m} = (x_1, \ldots, x_{m-1})$ be the context of level $m$ (empty for $m = 1$). Denote by $F_m(c_m)$ and $F'_m(c_m)$ the pre- and post-selection marginal frequencies of the context, and by $F(x_m \mid c_m)$ and $F'(x_m \mid c_m)$ the corresponding conditional distributions of level $m$. Define the forward and reverse divergences of level $m$ within context $c_m$,

$$D^{+}_{m,\boldsymbol{c_m}} = \sum_{x_m} F'(x_m|\boldsymbol{c_m}) \log \frac{F'(x_m|\boldsymbol{c_m})}{F(x_m|\boldsymbol{c_m})}, \quad D^{-}_{m,\boldsymbol{c_m}} = \sum_{x_m} F(x_m|\boldsymbol{c_m}) \log \frac{F(x_m|\boldsymbol{c_m})}{F'(x_m|\boldsymbol{c_m})}$$

and $J_{m,c_m} = D^{+}_{m,c_m} + D^{-}_{m,c_m}$ . Then the total Jeffreys information is

$$J[F',F] = \sum_{m=1}^{n} \sum_{\boldsymbol{c_m}} \left[ F'_m(\boldsymbol{c_m}) D^{+}_{m,\boldsymbol{c_m}} + F_m(\boldsymbol{c_m}) D^{-}_{m,\boldsymbol{c_m}} \right]$$

Equivalently,

$$J[F',F] = \sum_{m=1}^{n} \sum_{\boldsymbol{c_m}} \bar{F}_m(\boldsymbol{c_m}) J_{m,\boldsymbol{c_m}} + E_{\text{cpl}}, \quad \bar{F}_m(\boldsymbol{c_m}) = \frac{F_m(\boldsymbol{c_m}) + F'_m(\boldsymbol{c_m})}{2}$$

$$E_{\text{cpl}} = \frac{1}{2} \sum_{m=1}^{n} \sum_{\boldsymbol{c_m}} \left( F'_m(\boldsymbol{c_m}) - F_m(\boldsymbol{c_m}) \right) \left( D^{+}_{m,\boldsymbol{c_m}} - D^{-}_{m,\boldsymbol{c_m}} \right)$$

$E_{\text{cpl}}$ generalizes the between-level coupling term $E_{\text{SI}}$ of Section 4.4 ($E_{\text{GI}}$ in the group-class model above): for $n = 2$ it reduces to it. It should not be confused with the within-level residuals $E_{\text{holo}}$, $E_{\text{multi}}$ and $E_{\text{n}}$, which complete the partition of a local divergence into marginal and association terms.

### *Proof*

Write $\log(F'/F) = \Sigma_m \log[F'(x_m|c_m)/F(x_m|c_m)]$ and multiply by $F' - F$. The $m$-th term depends on $x$ only through $(c_m, x_m)$, so summing over the lower levels gives

$$\Sigma_{c_m,x_m}[F'm(c_m)F'(x_m|c_m)-F_m(c_m)F(x_m|c_m)]\log[F'(x_m|c_m)/F(x_m|c_m)]=$$

$$=\Sigma c_m[F'm(c_m)D^+m,c_m+F_m(c_m)D^-{}_{m,c_m}].$$ The second form follows from

$$F'D^+ + F D^- = \bar{F}(D^+ + D^-) + ½(F'-F)(D^+ - D^-).$$

For $m = 1$ the context is empty, $F_1 = F'_1 = 1$, and the level contributes $J_1$ with no coupling. For $n = 2$ the identity reduces to the one-level identity of the main text (Section 4.4).

*Consequences*

The identity shows how the local partitions combine into the global information. The marginal, association and residual terms of each level (Section 5 of the main text) partition the local divergence $J_{m,c_m}$, which is computed within a fixed context and remains valid as such. The global information $J[F', F]$, however, is not the sum of these local divergences: each $J_{m,c_m}$ enters weighted by the mean frequency $\bar{F}_m(c_m)$ of its context, because the context itself changes frequency under selection, and an additional coupling term $E_{\text{cpl}}$ is required. The coupling of a given level and context vanishes if the context frequency does not change ($F'_{m(c_m)} = F_{m(c_m)}$) or if the forward and reverse divergences coincide ($D^+_{m,c_m} = D^-_{m,c_m}$); it has no definite sign, and the total $E_{\text{cpl}}$ may also vanish by cancellation among levels and contexts. Under weak change $D^+ - D^-$ is of third order, so the between-level coupling $E_{\text{cpl}}$ is of higher order than the local Jeffreys terms. The complete hierarchical partition therefore consists of the local partitions of all levels, weighted as indicated, plus one coupling term for each level and context.